\pdfoutput=1
\documentclass[11pt,authoryear,a4paper]{elsarticle}
\usepackage[margin=2.5cm]{geometry}
\usepackage[latin1]{inputenc}
\usepackage{amsmath}
\usepackage{amsfonts}
\usepackage{amssymb}
\usepackage{amsthm}
\usepackage{natbib}
\usepackage{graphicx}
\usepackage{subfig}
\usepackage{url}
\usepackage{algorithm}
\usepackage{comment}
\usepackage{algpseudocode}

\theoremstyle{definition}

\begin{document}

\begin{frontmatter}

\title{Approximate maximum likelihood estimation using data-cloning ABC}

\author[lund]{Umberto~Picchini\corref{cor1}}
\ead{umberto@maths.lth.se}

\author[lund]{Rachele~Anderson}
\ead{rachele@maths.lth.se}

\cortext[cor1]{Corresponding author}

\address[lund]{Centre for Mathematical Sciences, Box 118 Lund University, SE-22100 Lund, Sweden}

\begin{abstract}

A maximum likelihood methodology for a general class of models is presented, using an approximate Bayesian computation (ABC) approach. The typical target of ABC methods are models with intractable likelihoods, and we combine an ABC-MCMC sampler with so-called ``data cloning'' for maximum likelihood estimation. Accuracy of ABC methods relies on the use of a small threshold value for comparing simulations from the model and observed data. The proposed methodology shows how to use large threshold values, while the number of data-clones is increased to ease convergence towards an approximate maximum likelihood estimate. We show how to exploit the methodology to reduce the number of iterations of a standard ABC-MCMC algorithm and therefore reduce the computational effort, while obtaining reasonable point estimates. Simulation studies show the good performance of our approach on models with intractable likelihoods such as $g$-and-$k$ distributions, stochastic differential equations and state-space models. 
\end{abstract}


\end{frontmatter}

\section{Introduction}

We present a methodology for approximate maximum likelihood estimation that uses approximate Bayesian computation (ABC, \citealp{tavare1997inferring}, \citealp{pritchard1999population}, \citealp{marjoram2003markov}). The method is applicable to a very general experimental setup, valid for both ``static'' and ``dynamic'' models. 
Our main question is: since in ABC studies artificial datasets are produced from the data generating model and compared to the observed data according to a threshold parameter $\delta$ (the smaller the $\delta$ the better the inference), what can we do if we are unable to reduce $\delta$ below a certain level? Or alternatively, can we perform inference using a relatively large $\delta$ and fewer Markov chain Monte Carlo (MCMC) simulations, instead of progressively decrease $\delta$ at the expense of using many MCMC iterations?
Suppose we have obtained a very rough approximation to the posterior distribution for the unknowns, if we have at least located the main mode for the posterior can we conduct approximate maximum likelihood inference? Basically by accepting of having obtained a poor approximation to the posterior, except for the location of its main mode, we switch to maximum likelihood estimation by proposing draws around such approximated mode using a special ABC-MCMC sampler.

Let $Y\sim f(y|\cdot,\phi)$ denote a realization from an observable random variable, i.e. $f(\cdot)$ is the data generating mechanism. Depending on the modelling scenario, $Y$ might be observed conditionally on an unobservable random variable $X$, i.e. $Y\sim f(y|X,\phi)$, or $X$ might be fixed and known (e.g. a set of deterministic inputs or covariates). In any case, there is dependence on an unknown (vector) parameter $\phi$. The statistical methods we are going to introduce have general appeal, however in order to set a working framework for the time being we assume to deal with state-space models (also known as Hidden Markov Models, \citealp{cappe2005inference}). We remark that our methods based on ABC are \textit{not} restricted to state-space, nor dynamic models. For example, below we assume $X$ a Markov process, but this is not a requirement nor is conditional independence of measurements.

Consider an observable, discrete-time stochastic process $\{Y_t\}_{t\geq t_0}$, $Y_t\in\mathsf{Y}\subseteq \mathbb{R}^{d_y}$ and a latent and unobserved continuous-time stochastic process $\{X_t\}_{t\geq t_0}$, $X_t\in\mathsf{X}\subseteq \mathbb{R}^{d_x}$.
Process $X_t\sim g(x_t|x_{t-1},\eta)$ is assumed Markov with transition densities $g(\cdot)$ depending on another unknown (vector) parameter $\eta$.
We think at $\{Y_t\}$ as a measurement-error-corrupted version of $\{X_t\}$ and assume that observations for $\{Y_t\}$ are conditionally independent given $\{X_t\}$.
Our state-space model can be summarised as
\begin{equation}
\begin{cases}
Y_t\sim f(y_t|X_t,\phi),\qquad t\geq t_0\\
X_t\sim g(x_t|x_{t-1},\eta).
\end{cases}
\label{eq:state-space-general}
\end{equation}
Typically $f(\cdot)$ is a known function set by the modeller, whereas $g(\cdot)$ is often unknown (e.g. when $\{X_t\}$ is a diffusion process, i.e. the solution of a stochastic differential equation) except for very simple toy models.
Goal of our work is to estimate the parameters $\theta=(\eta,\phi)$ using observations $y=(y_0,y_1,...,y_n)$ from $\{Y_t\}_{t\geq t_0}$ collected at discrete times $\{t_0,t_1,...,t_n\}$. 
As an example,  $\{Y_t\}_{t\geq t_0}$ may be defined as 
\begin{equation}
Y_{t_i}=X_{t_i} + \epsilon_{t_i}, \quad \epsilon_{t_i}\sim p_{\epsilon}(\phi), \qquad i=0,1,...,n \label{eq:measurement-model}
\end{equation}
with $\{\epsilon_t\}$ representing unobservable noise sources (e.g. measurement errors) having distribution with probability density function (pdf) $p_\epsilon(\cdot)$.

In Bayesian inference the goal is to analytically derive the posterior distribution $\pi(\theta|y)$ or, most frequently, implement an algorithm for sampling draws from the posterior distribution. Sampling procedures are often carried out using Markov chain Monte Carlo (MCMC) or Sequential Monte Carlo (SMC) methods embedded in MCMC procedures \citep{andrieu2010particle}. However the last ten years have seen an explosion of methodological advances for the so-called approximate Bayesian computational methods. We propose to use an ABC-MCMC sampler as a ``workhorse'' to obtain an approximate MLE for $\theta$. Notice that (approximate) Bayesian algorithms leading to an approximate MLE have also been considered in \cite{rubio2013simple}. See \cite{grazian-liseo(2015)} for ABC strategies for ``integrated likelihoods'', where nuisance parameters are integrated out.

The paper is organized as follows: in section \ref{sec:data-cloning} we briefly review some properties of ``data cloning'' for maximum likelihood estimation; in section \ref{sec:ABC-all} we introduce topics of approximate Bayesian computation methodology, first by considering some basics (section \ref{sec:abc-basics}) then introducing our original contribution in sections \ref{sec:abc-dc}--\ref{sec:dynamic-abcdc}. Finally in section \ref{sec:simulations} simulation studies illustrate results.

\subsection{Data cloning}\label{sec:data-cloning}

A Bayesian procedure for maximum likelihood estimation based on ``replicating'' data was first introduced in \cite{robert1993} (see also \citealp{robert-titterington}) and then studied under different flavors by e.g. \cite{doucet-godsill-robert}, \cite{jacquier-johannes-polson} and \cite{lele-dennis-lutscher}. The latter introduced the term ``data-cloning'' which we employ.
For simplicity, in the following we always consider the full vector parameter $\theta$, and it should be understood that some of its components enter $f(\cdot)$ while others enter $g(\cdot)$. 
The likelihood function of $\theta$ for the state-space model \eqref{eq:state-space-general} can be written as 
\begin{align}
L(\theta;y)&=p(y_0;\theta)\prod_{i=1}^np(y_i|y_1,...,y_{i-1};\theta)\nonumber\\
&=
\int f(y_0|x_0;\theta)p(x_0) \prod_{i=1}^n \bigl\{ f(y_i|x_i,\theta)g(x_i|x_{i-1},\theta)\bigr\}dx_0\cdots dx_n
\label{eq:likelihood}
\end{align} 
where $x_0=X_{t_0}$ and $p(x_0)$ the corresponding unconditional density. Actually in what follows, and without loss of generality, we assume $x_0$ deterministic and known, hence we remove $p(x_0)$ from the expression of the likelihood function.
Notice the latter equality in \eqref{eq:likelihood} exploits the notion of conditional independence between observations and the Markovian nature of the latent state. We now consider ``cloning'' the data $y$, i.e. we choose a positive integer $K$, produce $K$ copies of $y$ and stack them in $y^{(K)}=(y,y,...,y)$ where $y$ is replicated $K$ times. 
Now, we generate $K$ independent vectors for process $\{X_t\}_{t\geq t_0}$ from $g(\cdot)$ at times $\{t_0,...,t_n\}$, say $X^{(1)},X^{(2)},...,X^{(K)}$, all simulated conditionally on the same value of $\theta$. Therefore we set $X^{(k)}=(X_{t_0}^{(k)},...,X_{t_n}^{(k)}\}$ for a generic $k$ ($k=1,...,K$).
All vectors in $y^{(K)}$ are assumed to be conditionally independent of each other, given their individual latent state $X^{(k)}$, for example we could imagine a series of $K$ independent experiments leading to exactly the same result.  Then the likelihood function for the cloned data $y^{(K)}$ results in
\begin{equation}
L(\theta;y^{(K)})=\int \prod_{k=1}^K \{f(y|X^{(k)},\theta)p(X^{(k)}|\theta)\}dX^{(1)}\cdots dX^{(K)}
\label{eq:cloned-likelihood}
\end{equation}
where we denote with $p(X^{(k)}|\cdot)$ the joint density of vector $X^{(k)}$, hence $p(X^{(k)}|\theta)=\prod_{i=1}^n g(X^{(k)}_i|X^{(k)}_{i-1},\theta)$.
Now, for each $k$ a given term in the product in \eqref{eq:cloned-likelihood} depends on $X^{(k)}$ and not on terms having different indices $k'$ ($k'\neq k$). Therefore \eqref{eq:cloned-likelihood} is a product of $K$ integrals, each returning the likelihood function based on $y$, and we can write
  \begin{equation}
L(\theta;y^{(K)})=\prod_{k=1}^K \int f(y|X^{(k)},\theta)p(X^{(k)}|\theta)dX^{(k)} = (L(\theta;y))^K.
\label{eq:cloned-likelihood-bis}
\end{equation}
Therefore the likelihood function for the cloned data is the likelihood based on the actual measurements raised to the power $K$. Now, it is clear that the MLE of $\theta$ is the argmax for both $L(\theta;y^{(K)})$ and $L(\theta;y)$.
By considering a prior distribution $\pi(\theta)$, we have the posterior distribution resulting from a data-cloned likelihood
\begin{equation}
\pi(\theta|y^{(K)})\propto L(\theta;y^{(K)})\pi(\theta)=(L(\theta;y))^K\pi(\theta).
\label{eq:cloned-data-posterior}
\end{equation} 
It is easy to prove that for a large enough $K$ the mean of the posterior $\pi(\theta|y^{(K)})$ approaches the MLE of $\theta$ regardless the specific choice of $\pi(\theta)$ \citep{lele-dennis-lutscher} and a central limit theorem can be derived \citep{jacquier-johannes-polson,lele2010estimability}. However, simulations experiments in \cite{lele-dennis-lutscher} show that using informative priors enable a more rapid convergence to the MLE.\\

In algorithm \ref{alg:generalized-lele-algorithm} we consider a generalization of the data-cloning Metropolis-Hastings algorithm for sampling from $\pi(\theta|y^{(K)})$ (we call it ``generalization'' simply because in \citealp{lele-dennis-lutscher} the proposal distribution for $\theta$ is $\pi(\theta)$, while we consider a general proposal $u(\cdot)$).
Consider a proposed value $\theta^{\#}$ generated from a distribution having density $u(\theta^{\#}|\theta^*)$ and a proposal $X^{\#(k)}$ generated from $v(X^{\#(k)}|\theta^{\#})$ which we set for convenience to be $v(X^{\#(k)}|\theta^{\#})\equiv p(X^{\#(k)}|\theta^{\#})$.
\begin{algorithm}
\caption{A data-cloning MCMC algorithm}
\begin{algorithmic}
\State 1.  Initialization: Fix a starting value $\theta^*$ or generate it from its prior $\pi(\theta)$ and set $\theta_1=\theta^*$. Set $j=1$.
\State 2. Generate $K$ independent values of $X$, denoted $X^{*(1)},...,X^{*(K)}$ from $p(X|\theta^{*})$.
\State 3. Calculate 
\begin{equation*}
q^*=\prod_{k=1}^K f(y|X^{*(k)},\theta^*).
\end{equation*}
\State 4. Generate a $\theta^{\#}\sim u(\theta^{\#}|\theta^*)$. Generate independent $X^{\#(1)},...,X^{\#(K)}$ from $p(X|\theta^{\#})$. Compute $q^{\#}=\prod_{k=1}^K f(y|X^{\#(k)},\theta^{\#})$.
\State 5. Generate a uniform random variable $\omega\sim U(0,1)$, and calculate the acceptance probability 
\begin{eqnarray}
& \alpha = \min\biggl[1,\underbrace{\frac{q^{\#}\cdot p(X^{\#(1)}|\theta^{\#})\cdots p(X^{\#(K)}|\theta^{\#})}{q^* \cdot p(X^{*(1)}|\theta^{*})\cdots p(X^{*(K)|\theta^{*}})  }}_{\text{ratio of likelihoods}}\nonumber\\
&\times  \underbrace{\frac{v(X^{*(1)}|\theta^{*})\cdots v(X^{*(K)}|\theta^{*})u(\theta^*|\theta^{\#})}{v(X^{\#(1)}|\theta^{\#})\cdots v(X^{\#(K)}|\theta^{\#})u(\theta^{\#}|\theta^{*})}}_{\text{ratio of proposals}} \times \underbrace{\frac{\pi(\theta^{\#})}{\pi(\theta^*)}}_{\text{ratio of priors}} \biggr]\nonumber\\
&= \min\biggl[1,\frac{q^{\#}}{q^* }
\times  \frac{u(\theta^*|\theta^{\#})}{u(\theta^{\#}|\theta^{*})} \times \frac{\pi(\theta^{\#})}{\pi(\theta^*)} \biggr]. \label{eq:generalized-acceptance-prob}
\end{eqnarray}
If $\omega>\alpha$, set $\theta_{j+1}:=\theta_{j}$ otherwise set $\theta_{j+1}:=\theta^{\#}$, $\theta^*:=\theta^{\#}$ and $q^*:=q^{\#}$. Increase $j$ by 1 and go to step 6. 
\State 6. Repeat steps 4--5 as long as $j\leq R$ for $R$ ``large''.
\end{algorithmic}
\label{alg:generalized-lele-algorithm}
\end{algorithm}
The notation $:=$ means ``assign the value on the right hand side to the left hand side''. For a large enough number of iterations $R$ this algorithm produces a chain having $\pi(\theta,X|y^{(K)})$ as its stationary distribution, with $X=(X^{(1)},...,X^{(K)})$. In order to obtain draws from the desired marginal distribution $\pi(\theta|y^{(K)})$ it is sufficient to discard the $\{X^{(1)},...,X^{(K)}\}_{j=1,...,R}$ obtained from the algorithm output $\{\theta,X^{(1)},...,X^{(K)}\}_{j=1,...,R}$ (after some appropriate burnin period). For $K\rightarrow\infty$ the sample mean of the $\{\theta\}_j$ is the MLE of $\theta$ and $K$ times the covariance matrix of the draws returns the covariance of the MLE, the inverse of the Fisher information based on the original data  \citep{jacquier-johannes-polson,lele2010estimability}. Also,  for $K\rightarrow \infty$ and independently of the chosen prior, $\pi(\theta|y^{(K)})$ is degenerate at $\theta=\hat{\theta}$, where $\hat{\theta}$ is the MLE of $\theta$.

Notice the simplification occurring in the expression for $\alpha$, due to taking $v(X|\theta)\equiv p(X|\theta)$, this resulting in \eqref{eq:generalized-acceptance-prob}. The simplification above solves the typically difficult problem of not having a ready expression for the transition densities of $\{X_t\}_{t\geq t_0}$. In fact, here all we need is the ability to (somehow) simulate the process $\{X_t\}_{t\geq t_0}$, and having access to transition densities is not strictly required. For example, in section \ref{sec:gompertz} we know the solution of the considered stochastic differential equation (SDE) model, so we can simulate from it. When the exact solution to an SDE is not available, a numerical discretization with stepsize $h$ (e.g. the Euler-Maruyama scheme) generates an approximate solution, converging to the exact one as $h\rightarrow 0$. \cite{beskos2006exact} even devised a numerical scheme resulting in exact simulation of the SDE solution (i.e. without discretization error), though this is of not so general applicability.

It is important to realize that dealing with a ``powered-up'' posterior as in \eqref{eq:cloned-data-posterior} results in a surface having increasingly peaked modes for increasing $K$ and deeper ``valleys'' in-between modes enclosing smaller and smaller probability mass (assuming the existence of multiple modes). Therefore we believe it is important not to let $K$ fixed to a large value from the start of the algorithm, but instead start with a small value for $K$ and then increase it progressively. Enabling a smooth and not too rapid increase of $K$ should help the chain from being stuck in low-probability regions. However in the examples discussed in sections \ref{sec:g-and-k}--\ref{sec:2GBM} a rapid increase in $K$ is possible.

\section{Approximate inference using ABC with data-cloning}\label{sec:ABC-all}

Acceptance of proposals in MCMC algorithms is particularly challenging when the modelled process is highly erratic, for example when the unobserved state is a diffusion process, that is a solution to a stochastic differential equation (SDE, e.g. \citealp{fuchs2013inference}). For such class of models, trajectories for $\{X_t\}$ may result quite distant from the observed data $y$, even for values of the parameters in the bulk of their posterior distributions. In such circumstance $q^{\#}$ will often be small compared to $q^{*}$, and the proposal will rarely be accepted. For example, when using an approach as the one described above, where trajectories are simulated ``blindly'' from $p(X|\cdot)$, that is unconditionally to data, then trajectories do not exploit direct knowledge of the data. This sometimes result in many rejected proposals if the sample size is large. Carefully tuned Sequential Monte Carlo (SMC) strategies can be constructed so that the best trajectories (``particles'') are selected according to their proximity to data, and this have pushed forward Bayesian inference via MCMC methods incorporating SMC \citep{andrieu2010particle}. 

However for complex (ideally multidimensional) stochastic models and a large number of observations, use of SMC methods is computer intensive. Approximate Bayesian computation (ABC, see reviews by \citealp{sisson-fan(2011)} and \citealp{marin-et-al(2011)}) eases sampling from an approximation of the posterior distribution,  by substituting likelihood function evaluations with simulations from the data generating model. Here follows a short discussion on ABC which will ease the introduction to our original contribution. 

\subsection{Basics of ABC}\label{sec:abc-basics}

Here we summarize some minimal notions of ABC methodology, without considering for the moment the data-cloning scenario, hence in this section it can be assumed that $K=1$.
The ABC approach considers generating samples $z$ from $f(\cdot)$ in \eqref{eq:state-space-general} (i.e. $z\in \mathsf{Y}$, same as the actual data) and corresponding proposals $\theta^{\#}$ are accepted if the $z$ are ``close'' to data $y$, according to a threshold $\delta>0$. Several criterion for ``closeness'' can be postulated, as described below. In ABC we aim at simulating draws from the augmented approximated posterior
\begin{equation}
\pi_{\delta}(\theta,z|y)\propto J_{\delta}(y,z)\underbrace{L(\theta;z)\pi(\theta)}_{\propto \pi(\theta|z)}
\label{eq:abc-posterior}
\end{equation}
where $z=(z_0,...,z_n)$ and $L(\theta;z)$ is the (intractable) likelihood function for $\theta$ based on $z$. Then $\pi_{\delta}(\theta|y)\propto \int \pi_{\delta}(\theta,z|y)dz$. Here $J_{\delta}(\cdot)$ is some function that depends on $\delta$ and weights the intractable posterior for simulated data $\pi(\theta|z)\propto L(\theta;z)\pi(\theta)$ with high values in regions where $z$ and $y$ are similar. Therefore we would like (i) $J_{\delta}(\cdot)$ to give higher rewards to proposals corresponding to $z$ having values close to $y$. In addition (ii) $J_{\delta}(y,z)$ is assumed to be a constant when $z=y$ (i.e. when $\delta=0$) so that the \textit{exact} marginal posterior $\pi(\theta|y)$ is recovered. 
A common choice for $J_{\delta}(y,z)$ is the uniform kernel
\[J_{\delta}(y,z)\propto \mathbb{I}_{\{\rho(z,y)\leq\delta\}}\]
where $\rho(z,y)$ is some measure of closeness between $y$ and $z$ and $\mathbb{I}$ is the indicator function. Important alternatives are the Epanechnikov and Gaussian kernels \citep{beaumont2010approximate} or sums of discrepancies \citep{toni2009approximate}. 
An ABC-MCMC algorithm targeting the distribution \eqref{eq:abc-posterior} has been proposed in \cite{marjoram2003markov}. However, one of the difficulties is that, in practice, $\delta$ has to be set as a tradeoff between statistical accuracy (with a small positive $\delta$) and computational feasibility ($\delta$ not too small). Also, notice that ABC methods are most often applied to models where a set of low-dimensional summaries of the data $S(y)$ is employed rather than the full dataset $y$. That is whenever is possible (and even more so when $S(\cdot)$ is sufficient for $\theta$), it is advisable to consider $J_{\delta}(S(y),S(z))$ instead, so that for example we have 
\[
\pi_{\delta}(\theta,z|\rho(S(z),S(y))\leq\delta)\propto L(\theta;z)\pi(\theta)\mathbb{I}_{\{\rho(S(z),S(y))\leq\delta\}}
\]
when $J_{\delta}(S(y),S(z))\propto \mathbb{I}_{\{\rho(S(z),S(y))\leq\delta\}}$.

The introduction of summaries $S(\cdot)$ is a double edged sword. On one hand the specification of appropriate (i.e. informative, though usually not sufficient) statistics is not trivial, especially for dynamic models, whereas it is somehow more intuitive to specify them for static models. On the other hand having an informative set of statistics implies a considerable reduction in the number of elements to be compared ($d_s$ comparisons, where $d_s:=\dim(S)$ instead of the $n$ comparisons required when simulated and observed data have to be compared, with $d_s\ll n$) and consequently a much smaller $\delta$ can be employed. 

\subsection{ABC-DC: Data-cloning ABC}\label{sec:abc-dc}

In most problems $\delta$ is a strictly positive value, sometimes set ``small enough'', some other times set to a larger than desired value, depending on the complexity of the experimental scenario. In fact when using a very small $\delta$ to obtain accurate inference, this results in a high rejection rate, often too high to be computationally feasible. Therefore we propose to consider a larger $\delta$ than what would typically be considered as appropriate, coupled with data-cloning. Our idea is that if we use $K=1$ while dynamically decrease $\delta$ in an ABC-MCMC algorithm, to reach a moderately large $\delta$-value that still allows exploration of the posterior surface (producing an acceptance rate of, say, 10-15\%, and this phase could be considered as ``burn-in'') we can then start a data-cloning procedure and progressively enlarge $K$ while keeping $\delta$ constant to its last value. During the initial exploration (burn-in with $K=1$) we require a $\delta$ which is small enough to locate the approximate position of the maxima for the exact marginal posterior $\pi(\theta|y)$, not a $\delta$ producing an accurate approximation to the surface of $\pi(\theta|y)$. When we start increasing $K$ the marginal posterior $\pi_{\delta}(\theta|y^{(K)})$ will concentrate around its maxima, which for large enough $K$ should be approximately at the same location as the MLE.

More in detail we propose to consider $R$ iterations of an ABC-MCMC algorithm later denoted as ABC-DC: (i) start an ABC-MCMC algorithm without data-cloning ($K=1$) and let $\delta$ decrease during this phase; an initially large $\delta$ will enable a rapid exploration of the posterior surface at a high acceptance rate (say 30\%) to locate the bulk of the approximated posterior. (ii) while $\delta$ progressively decreases, the algorithm focus on exploring a more accurate approximation of the posterior until $\delta$ reaches say 10-20\% acceptance rate. Such acceptance rate is typically too high  in ABC studies (a typical value would be 1\% or less) however we plan to increase the number of clones and focus on the peak of such posterior. (iii) At this point $\delta$  is kept fixed to its last value and data-cloning starts, by progressively increasing the value of $K$. (iv) Once $R$ iterations are completed, we collect the draws generated with the largest value of $K$ and use those for maximum likelihood inference. As such the resulting samples are not from $\pi_{\delta}(\theta,z|y)$ but from the powered-approximated posterior $\pi_{\delta}(\theta,z^{(1)},...,z^{(K)}|y^{(K)})$ for finite $K$
\begin{align*}
\pi_{\delta}(\theta,z^{(1)},...,z^{(K)}|y^{(K)}) \propto \pi(\theta)\prod_{k=1}^K J_\delta(y,z^{(k)})L(\theta;z^{(k)})
\end{align*}
where the $L(\theta;z^{(k)})$ can be simplified out in the Metropolis-Hastings acceptance probability as previously illustrated. As mentioned in section \ref{sec:abc-basics}, we  assume as implicit the dependence on data via summary statistics, therefore here and in the rest of our work $J_\delta(y,z^{(k)})\equiv J_\delta(S(y),S(z^{(k)}))$.

Of course enlarging $K$ shrinks the area of the support of the cloned posterior where most of the probability mass is located, hence it becomes increasingly difficult to explore a progressively peaked surface: this is why in step (ii) of the schedule above we do not recommend to go below, say, 10-20\% acceptance as this rate will reduce drastically when $K$ increases in step (iii).  

In our applications we use a Gaussian kernel, that is
\begin{equation}
J_{\delta}(y,z^{(k)}) \propto \exp\{-\bigl(S(z^{(k)})-S(y)\bigr)'\Omega^{-1}\bigl(S(z^{(k)})-S(y)\bigr)/2\delta^2\}
\label{eq:gauss-kernel}
\end{equation}
where $'$ denotes transposition.
Clearly equation \eqref{eq:gauss-kernel} respects desired criteria, i.e. (i) it is constant when $S(z^{(k)})\equiv S(y)$  and (ii) it gets larger values when  $S(z^{(k)})\approx S(y)$. For a generic $z$ this implies writing $S(z)\sim N_{d_s}(S(y),\delta^2\Omega)$ with $N_{d_s}(\cdot,\cdot)$ a $d_s$-dimensional Gaussian distribution centred at $S(y)$ and $\Omega$ a positive definite matrix. 
For simplicity we assume a diagonal $\Omega$ with elements $\Omega=\mathrm{diag}\{\omega^{2}_1,...,\omega^{2}_{d_s}\}$. Of course, using a diagonal $\Omega$ might have an impact on the inference as it does not take into account the correlation among summary statistics. Recall that we keep writing $J_{\delta}(y,z)$ instead of $J_{\delta}(S(y),S(z))$, the dependence on data via summary statistics being considered as implicit.

When the elements in vector $S(\cdot)$ are  varying approximately on the same range of values it is possible to consider $(\omega^{2}_1,...,\omega^{2}_{d_s})=(1,...,1)$, however in general the variability of the statistics is unknown and, depending on the type of data and the underlying model, these can have very different magnitude. An inappropriate choice for the elements in $\Omega$ affects the accuracy of the ABC inference negatively, with $J_{\delta}(\cdot)$ being dominated by the most variable
statistic so that $\delta$ will bound the distance with respect to such statistic, not the remaining ones.  
To our knowledge the only systematic study on the weighting of summary statistics in ABC is \cite{prangle-adaptiveABC}. When using summary statistics in our experiments, before starting the data cloning procedure we run a pilot study using ABC-MCMC (i.e. $K=1$) with $(\omega_1,...,\omega_{d_s})=(1,...,1)$ and collect the  values of the accepted $S(z)$. At the end of the pilot run we compute (after some appropriate burnin) the mean or the median absolute deviation MAD for each coordinate of the accepted $S(z)$ and define $(\omega_1,...,\omega_{d_s}):=(\mathrm{MAD}_1,...,\mathrm{MAD}_{d_s})$. Then we plug the obtained $\Omega$ to weight the summary statistics into data-cloning ABC (introduced in sections \ref{sec:abc-dc}--\ref{sec:dynamic-abcdc})  or in a further run of ABC-MCMC, for comparison purposes. See \cite{prangle-adaptiveABC} for a thorough study on this approach.

For ease of reading we produce two ABC-DC algorithms: the ``static'' ABC-DC is given in algorithm \ref{alg:abc-dc}, where both $\delta$ and $K$ are assumed fixed, allowing for a more immediate understanding. However, in our applications we use algorithm \ref{alg:abc-dc-dynam} which is discussed later. 

\begin{algorithm}
\caption{Static ABC-DC}
\begin{algorithmic}
\State 1. Initialization: Fix a starting value $\theta^*$ or generate it from its prior $\pi(\theta)$ and set $\theta_1=\theta^*$. Set $j=1$, fix $\delta>0$ and a positive integer $K$. A vector of statistics $S(\cdot)$ and weights $\Omega$ is available.
\State 2. Generate $K$ independent values of the latent process $X^{*(1)},...,X^{*(K)}$ from $p(X|\theta^{*})$. Conditionally on each $X^{*(k)}$ generate a corresponding $z^{*(k)}$ from equation \eqref{eq:state-space-general} for each $k=1,...,K$.
\State 3. Calculate $J_{\delta}(y,z^{*(k)})$ for every $k$ and compute
\begin{equation*}
q^*=\prod_{k=1}^K J_{\delta}(y,z^{*(k)}).
\end{equation*}
\State 4. Generate a $\theta^{\#}\sim u(\theta^{\#}|\theta^*)$. Generate $K$ independent $X^{\#(k)}$'s from $p(X|\theta^{\#})$ and corresponding $z^{\#(k)}$. 
\State 5. Calculate $J_{\delta}(y,z^{\#(k)})$ for every $k$ and set $q^{\#}=\prod_{k=1}^K J_{\delta}(y,z^{\#(k)})$.
Generate $\omega\sim U(0,1)$, and calculate  
\[
\alpha = \min\biggl[1,\frac{q^{\#}}{q^* }
\times  \frac{u(\theta^*|\theta^{\#})}{u(\theta^{\#}|\theta^{*})} \times \frac{\pi(\theta^{\#})}{\pi(\theta^*)} \biggr]. 
\]
If $\omega>\alpha$, set $\theta_{j+1}:=\theta_{j}$ and, increase $j$ by 1 and go to step 6. Otherwise set $\theta_{j+1}:=\theta^{\#}$, $\theta^*:=\theta^{\#}$, $q^*:=q^{\#}$, increase $j$ by 1 and go to 6. 
\State 6. Repeat steps 4--5 as long as $j\leq R$.
\end{algorithmic}
\label{alg:abc-dc}
\end{algorithm}

\subsection{Dynamic ABC-DC}\label{sec:dynamic-abcdc}

In our experiments, and unlike in \cite{lele-dennis-lutscher} and \cite{jacquier-johannes-polson} where $K$ is kept fixed during the MCMC algorithm execution, we let $K$ increase (see also \citealp{doucet-godsill-robert}). A ``dynamic'' version of ABC-DC with varying $\delta$ and $K$ is presented in algorithm \ref{alg:abc-dc-dynam}. Notice that previously cited work did not use ABC with data-cloning, so to the best of our knowledge ours is the first work proposing doing so.

As discussed by Christian P. Robert at \url{http://xianblog.wordpress.com/2010/09/22/feedback-on-data-cloning/} data-cloning share features with the simulated annealing global optimization method: keeping $K$ fixed to a high value once and for all
removes the dynamic features of a simulated annealing random walk that first explores the whole space and then progressively focus on the highest modes, achieving convergence if the cooling is slow enough. In other words, if $K$ is ``large enough'', the Metropolis algorithm will face difficulties in the exploration of the parameter space, and hence in the subsequent discovery of the global modes, while, if $K$ is ``too small'', there is no certainty that the algorithm will identify the right mode of possibly multiple modes. Here follow a ``dynamic'' ABC-DC algorithm, where schedules are defined for $\delta$ and $K$, namely $\{\delta_{r_1},\delta_{r_2},...,\delta_{r_p}\}$ and $\{ K_{s_1},K_{s_2},...,K_{s_q}\}$ with $\delta_{r_1}>...>\delta_{r_p}>0$ and $1=K_{s_1}<...<K_{s_q}$. This version of ABC-DC starts with $s_1$ iterations of ABC-MCMC algorithm with decreasing thresholds, where $s_1=\sum_{l=1}^pr_l$, $K=1$ constantly throughout the $s_1$ iterations. The threshold is $\delta:=\delta_{r_1}$ in the first $r_1$ iterations, $\delta:=\delta_{r_2}$ in the next $r_2$ iterations etc. In summary the first $r_1$ iterations use $(\delta,K):=(\delta_{r_1},1)$ and in general during iterations $(r_l:r_{l+1})$ we use $(\delta,K):=(\delta_{r_l},1)$. During the last $r_p$ iterations of ABC-MCMC we keep track of the maximum value $\mathrm{max}\{\pi_{\delta_{r_p}}(\theta|y)\}$ of the approximated posterior $\pi_{\delta_{r_p}}(\theta|y)$ and corresponding $\tilde{\theta}:=\mathrm{argmax}\pi_{\delta_{r_p}}(\theta|y)$: this is easily accomplished and cheap to implement by initializing $\mathrm{max}\{\pi_{\delta_{r_p}}(\theta|y)\}:=0$ just before setting $(\delta,K):=(\delta_{r_p},1)$. Then, whenever we have $J_{\delta}(y,z^{\#})\pi(\theta^{\#})>\mathrm{max}\pi_{\delta_{r_p}}(\theta|y)$ for the current maximised value of the posterior kernel, we set $\mathrm{max}\{\pi_{\delta_{r_p}}(\theta|y)\}:=J_{\delta}(y,z^{\#})\pi(\theta^{\#})$  and $\tilde{\theta}:=\theta^{\#}$. This search for the maximum has to be performed only when $\delta\equiv\delta_{r_p}$ as we are not interested in the maximum obtained for poorer approximations of the posterior. An alternative approach to the search for a mode $\tilde{\theta}$ is given by adjusting the output of ABC-MCMC as from \cite{beaumont2002approximate}, hence step $5'$ denoted as ``optional'' in algorithm \ref{alg:abc-dc-dynam}; this step is described in section \ref{sec:regression-adjustment}.

\begin{algorithm}
\caption{Dynamic ABC-DC}
\small
\begin{algorithmic}
\State \textbf{ABC-MCMC stage}
\State 1. Initialization: Fix a starting value $\theta^*$ or generate it from its prior $\pi(\theta)$ and set $\theta_1=\theta^*$. Set $j:=1$, $\delta:=\delta_{r_1}$ and  $\mathrm{max}\pi_{\delta_{r_p}}:=0$.
\State 2. Generate $X^{*}$ from $p(X|\theta^{*})$ and a corresponding $z^{*}$ from \eqref{eq:state-space-general}. Compute
$q^*=J_{\delta}(y,z^{*})$.
\State 3. Generate $\theta^{\#}:= \texttt{AMRW}(\theta^*,\Sigma_j)$. Generate $X^{\#}$'s from $p(X|\theta^{\#})$ and corresponding $z^{\#}$. Compute $q^{\#}=J_{\delta}(y,z^{\#})$.
\State 4.  Generate $\omega\sim U(0,1)$, and calculate 
\[
\alpha = \min\biggl[1,\frac{q^{\#}}{q^* } \times \frac{u_1(\theta^*|\theta^{\#},\Sigma_j)}{u_1(\theta^{\#}|\theta^{*},\Sigma_j)} 
\times \frac{\pi(\theta^{\#})}{\pi(\theta^*)} \biggr]. 
\]
If $\omega>\alpha$, set $\theta_{j+1}:=\theta_{j}$ otherwise set $\theta_{j+1}:=\theta^{\#}$, $\theta^*=\theta^{\#}$ and $q^*:=q^{\#}$. If $\delta=\delta_{r_p}$ go to 5. Otherwise increase $j$ by 1 and if $j\in \{r_2,...,r_p\}$ update $\delta:=\delta_j$.  Then go to 3.
\State 5. Check current maximum (only when $\delta=\delta_{r_p}$): if $q^{\#}\pi(\theta^{\#})>\mathrm{max}\pi_{\delta_{r_p}}$ set $\mathrm{max}\pi_{\delta_{r_p}}:=q^{\#}\pi(\theta^{\#})$ and $\tilde{\theta}:=\theta^{\#}$. Increase $j$ by 1. If $j\leq s_1$ go to 3 else go to step 6.
\State $5'.$ \textit{(optional)} Apply regression adjustment on the draws obtained with $\delta_{r_p}$. Take as $\tilde{\theta}$ either the mean or the mode of the adjusted draws and call $\Sigma_{s_1}$ the covariance of the adjusted draws.
\State
\State \textbf{Data-cloning stage}
\State 6. Take the last accepted value $\theta^*$, $\tilde{\theta}$ and the current covariance $\Sigma_{s_1}$ from ABC-MCMC. Set $\hat{\Sigma}_k:=\Sigma_{s_1}$, $K:=K_{s_2}$ and $\delta:=\delta_{r_p}$. 
\State 7. Generate $K$ independent vectors denoted $X^{*(1)},...,X^{*(K)}$ from $p(X|\theta^{*})$. Conditionally on each $X^{*(k)}$ generate a corresponding $z^{*(k)}$. Calculate
$q^*=\prod_{k=1}^K J_{\delta}(y,z^{*(k)}).$
\State 8. Generate $\theta^{\#}:=\texttt{MIS}(\tilde{\theta},\hat{\Sigma}_k)$. Generate $K$ independent $X^{\#(k)}$'s from $p(X|\theta^{\#})$ and corresponding $z^{\#(k)}$. Compute $q^{\#}=\prod_{k=1}^K J_{\delta}(y,z^{\#(k)})$.
\State 9. 
Generate $\omega\sim U(0,1)$ and calculate  
\[
\alpha = \min\biggl[1,\frac{q^{\#}}{q^* }
\times  \frac{u_2(\theta^*|\tilde{\theta},\hat{\Sigma}_k)}{u_2(\theta^{\#}|\tilde{\theta},\hat{\Sigma}_k)} \times \frac{\pi(\theta^{\#})}{\pi(\theta^*)} \biggr]. 
\]
If $\omega>\alpha$, set $\theta_{j+1}:=\theta_{j}$ otherwise set $\theta_{j+1}:=\theta^{\#}$, $\theta^*:=\theta^{\#}$ and $q^*:=q^{\#}$. 
Increase $j$ by 1. If $j\in \{s_{3},...,s_{q}\}$ increase $K$ and update $\hat{\Sigma}_k:=\hat{cov}(\theta)_{k'}$ then go to 7, otherwise go to 10. Here $\hat{cov}(\theta)_{k'}$ is the sample covariance computed on draws generated using the previous value $k'$ of $K$ in the schedule.
\State 10. If $j\leq R$ go to 8 otherwise stop.

\end{algorithmic}
\label{alg:abc-dc-dynam}
\end{algorithm}

During ABC-MCMC we generate parameter proposals using adaptive Gaussian Metropolis random walk \citep{haario-saksman-tamminen}. We write $\theta^{\#}:= \texttt{AMRW}(\theta^*,\Sigma_j)$ to denote such a proposal (and the corresponding Gaussian proposal function is denoted $u_1(\cdot)$). During the exploration of the approximate posterior surface with $K=1$ we aim at locating the principal mode $\tilde{\theta}$ of the distribution for the smallest threshold $\delta_{r_p}$ as described above. At the end of the $s_1$ iterations of ABC-MCMC we fix $\delta:=\delta_{r_p}$ for the rest of the ABC-DC execution and increase $K$ progressively. At this stage parameters are proposed using a Metropolis independent sampler $\texttt{MIS}(\tilde{\theta},\hat{\Sigma}_k)$ generating from the Gaussian distribution $N(\tilde{\theta},\hat{\Sigma}_k)$ (and the corresponding Gaussian proposal function is denoted $u_2(\cdot)$) where $\hat{\Sigma}_k$ is the sample covariance matrix obtained from draws generated using the previous value of $K$. 

We switched from a Gaussian random walk to an independent Metropolis sampler when powering-up because ABC-MCMC should have located the bulk of the approximated posterior (optionally with the help of the regression adjustment in section \ref{sec:regression-adjustment}), as well as its highest mode, and therefore we use such information to propose parameters in the next stage corresponding to a larger $K$. Not following random walk dynamics helps in avoiding getting trapped in local modes that might emerge when powering-up the posterior surface for increasing $K$. 
This is why, in order to accommodate the reduced support of the current targeted distribution for increasing $K$, we recompute the covariance matrix $\hat{\Sigma}_{k}$ for the independence sampler at iterations $j\in\{s_1,...,s_q\}$.

Notice that at the end of step 9 when we reach an iteration $j\in\{s_1,...,s_q\}$ and $K$ has to be enlarged, we need to ``balance'' the information contained in the numerator of $\alpha$ with the one in the denominator, and this is why we go back to step 7 instead of 8 and recompute $q^*$. Basically after increasing $K$ the number of clones in the numerator would be larger than the one in the denominator, hence the need to first go back to 7, where we recompute the denominator using the same value of $K$ employed for the numerator. This re-evaluation is performed only when $j\in \{s_1,...,s_q\}$ and we can safely interpret this step as the starting point for a new chain, targeting the corresponding powered distribution. The effect of step 7 is removed after some burnin period (and a presumably short one, as at this point the chain should already be in the bulk of the powered posterior). In the end, from the inference point of view, what matters are the draws generated at the largest $K$, which are generated with a fixed $\delta\equiv\delta_{r_p}$, hence are genuinely distributed according to the corresponding (marginal) posterior $\pi_\delta(\theta|y^{(K)})$.

For complex models a slow transition between different number of clones $K_{s_{j'}}$ might be required, meaning that differences $K_{s_{j'}}-K_{s_{j'+1}}$ should not be too large. This is because the chain needs to adapt to a narrower support for the posterior when using $K_{s_{j'+1}}$ clones, and to ease the exploration we use the covariance obtained from draws generated with $K_{s_{j'}}$ clones, which is hopefully appropriate. Otherwise if the difference above is too large the covariance used will be inappropriate (too large variances) and many proposals will be rejected.

A maximum likelihood ABC algorithm has also been proposed in \cite{rubio2013simple}, however in that work a non-parametric kernel estimator of the posterior density is constructed (using draws from the ABC rejection algorithm proposed in \citealp{pritchard1999population}), then the maximizer of the non-parametric density is found numerically. In our approach we do not require any kernel estimation procedure (which is onerous unless the dimension of $\theta$ is low), nor direct optimization procedures. An approach similar to the one by Rubio and Johansen is in \cite{grazian-liseo(2015)}.

In conclusion, we use draws produced by algorithm \ref{alg:abc-dc-dynam} under $(\delta,K)\equiv(\delta_{r_p},K_{s_q})$ and for these draws we compute their sample mean $\hat{\theta}_{\delta,K}$ to obtain an approximate maximum likelihood estimate (MLE) of $\theta$. In principle, if we were able to construct sufficient summaries $S(\cdot)$ for $\theta$, and by letting $\delta\rightarrow 0$ and then $K\rightarrow\infty$, we would have  $\hat{\theta}_{MLE}\equiv \hat{\theta}_{\delta,K}$ and $\hat{\Sigma}_{MLE} \equiv K\cdot\hat{\Sigma}_{\delta,K}$, where $\hat{\theta}_{MLE}$ is the MLE of $\theta$ and $\Sigma_{MLE}$ is the covariance of the MLE (i.e. the inverse of the Fisher information based on $y$) computed using the sample covariance $\hat{\Sigma}_{\delta,K}$ derived under $(\delta,K)\equiv(\delta_{r_p},K_{s_q})$. The reasoning behind these results, when $S(\cdot)$ is sufficient, is that (i) for an ABC approximation to a posterior it holds that in distribution $\lim_{\delta\rightarrow 0}\pi_{\delta}(\theta|y)=\pi(\theta|y)$, and that (b) data-cloning implies that in distribution $\lim_{K\rightarrow\infty}\pi(\theta|y^{(K)})=N(\hat{\theta}_{MLE},\hat{\Sigma}_{MLE})$ (\citealp{jacquier-johannes-polson}, \citealp{lele2010estimability}). Therefore by first taking the limit for $\delta$ and then applying the limit for $K$ we have that 
\[\hat{\theta}_{\delta,K}\sim N(\hat{\theta}_{MLE},\hat{\Sigma}_{MLE}),\qquad \delta\rightarrow 0, K\rightarrow \infty.\] 
Then (when asymptotics holds) it would be easy to compute approximate standard errors and construct confidence intervals for the true value $\theta^o$ of $\theta$ (using the fact that $\hat{\theta}_{MLE}\rightarrow \theta^o$ when $n\rightarrow\infty$), by noting that $\hat{\Sigma}_{MLE}$ is provided ``for free'' as detailed above. However, in reality our reasoning is based on not using a small $\delta$, therefore the confidence bounds resulting from using asymptotics are often wide, even though in our experiments we obtain good point approximations to the MLE. This is because $\delta$ is large and an increasing $K$ does not necessarily reduce the chain variability for all parameters. Therefore multiplying $\hat{\Sigma}_{\delta,K}$ by $K$ might give too large standard errors (this fact holds regardless of whether we are able to make use of sufficient summary statistics, which is in general not the case). This is particularly true when summary statistics are not informative for a certain parameter, see the case of parameter $\log\sigma$ in Figure \ref{fig:gompertz-regularization-top}. For all these reasons, focus of this work is on parameters point estimation.

In conclusion, even if we use a $\delta$ which is not small for accurate Bayesian inference, but small enough for locating the maximum of the posterior, then we can power-up the ABC posterior and propose samples around such maximum to obtain a good (point) approximation of the MLE. A formal study on the properties of the obtained estimators for positive $\delta$ and finite $K$ is not considered here and is left for future research.

Finally note that in \cite{lele2010estimability} it is shown that a criterion to choose the number of clones is to monitor the decay (as $K$ increases) of the largest eigenvalue $\lambda_1(K)$ of the sample covariance matrix obtained from a chain using $K$ clones. They used such criterion to diagnose parameters estimability, that is if $\lambda_1(K)$ decreases to zero then the parameters are deemed estimable, with a covariance matrix approaching degeneracy. The examples treated with standard data-cloning (i.e. papers not using ABC methodology) are sometimes able to consider clones in the order of hundreds \citep{baghishani-mohammadzadeh} with a good acceptance rate, hence it makes sense to monitor the convergence of $\lambda_1(K)$. When using ABC we do not enjoy the luxury of letting an automatic criterion decide when to stop increasing $K$, as we are anyway bounded to use a much smaller number of clones with a small acceptance rate.

\subsection{Regression adjustment}\label{sec:regression-adjustment}

In section \ref{sec:g-and-k} we consider an example with a static model, where ABC inference is easily enabled thanks to  the existence of intuitive and informative summary statistics (albeit not sufficient ones). However for dynamic models it is usually way more difficult to identify informative summaries. Therefore in section \ref{sec:gompertz} we automatically retrieve summaries using the method in \cite{fearnhead-prangle(2011)}. Although automatized construction of summaries is certainly a handy tool, since we plan to use a large threshold $\delta$ at the end of the burnin ($K=1$) phase it may be useful to ``adjust'' the obtained draws before starting the data-cloning phase, in order to obtain a more informative independence sampler. For the example in section \ref{sec:gompertz} we consider the regression adjustment proposed in \cite{beaumont2002approximate}. In this section for ease of writing we assume a scalar $\theta$.
Denote with $\theta_{\delta}=(\theta_1,...,\theta_{r_p})$ the sequence of draws for $\theta$ produced at the smallest threshold $\delta\equiv\delta_{r_p}$ when $K=1$ and denote with $(S_1,...,S_{r_p})$ the corresponding simulated summary statistics (each summary can be a vector). Consider the following regression model
\begin{equation}
\theta_i = \alpha + (S_i-S)'{\beta}+\xi_i,\qquad i=1,...,r_p
\label{eq:beaumont-regression}
\end{equation}
where $S$ is the summary statistic for the observed data, $\alpha$ and $\beta$ are regression parameters and $\xi_i$ is mean zero homoscedastic noise (see \citealp{blum2013comparative} for alternative approaches). Parameters $(\alpha,\beta)$ can be estimated via local linear regression by minimizing the following criterion 
\[
\sum_{i=1}^{r_p}(\theta_i-\alpha-(S_i-S)'\beta)^2 J_{\delta_{r_p}}(S,S_i)
\]
for some appropriate kernel $J_{\delta}$ (e.g. uniform, Gaussian kernel). Solution to the least squares problem is given by
\[
(\hat{\alpha},\hat{\beta}) = (Z^TWZ)^{-1}Z^TW\theta_{                        \delta}
\]
where $Z$ is the design matrix for model \eqref{eq:beaumont-regression} and $W$ a diagonal matrix with $i$th entry given by $J_{\delta_{r_p}}(S,S_i)$ (see \citealp{beaumont2002approximate} for details). The adjusted parameters are given by
\[
\theta_i^* = \theta_i - (S_i-S)^T\hat{\beta},\qquad i=1,...,r_p.
\]
When the employed $\delta_{r_p}$ is relatively large the posterior obtained from the adjusted draws $\theta_i^*$ is usually more concentrated than the one based on $\theta_i$, see section \ref{sec:gompertz}. This means that, based on the set of $\theta_i^*$, we are able to construct a more informative empirical covariance matrix for the independence sampler used when $K>1$. Also, we may consider taking the mean or median of the adjusted parameters and centre the independence sampler at such value ($\tilde{\theta}$ in algorithm \ref{alg:abc-dc-dynam}).

\section{Simulation studies}\label{sec:simulations}

This section considers approximate inference for three simulation studies. The first two studies deal with observations from a $g$-and-$k$ distribution and from a state space model respectively, both lacking explicit expressions for the likelihood function. The third one is based on a two-dimensional stochastic differential equation with correlated noise. For the latter it is possible to write the exact likelihood function and therefore identify by numerical optimization the maximum likelihood estimate, which we compare with the ABC-DC estimator. In all examples whenever we refer to ABC-DC we mean the ``dynamic ABC-DC'' in algorithm \ref{alg:abc-dc-dynam}.

From the computer coding point of view, the three examples are easily vectorised (our codes are written in MATLAB) and therefore simulating model realizations for a certain $K>1$, say $5\leq K \leq 15$, did not result in any serious slow-down compared to using $K< 5$. However, consider having a computationally expensive model simulator such that producing a single realizations from the model requires several seconds or minutes. Assume therefore that running many ($R$) iterations of an ABC-MCMC algorithm for Bayesian inference is impractical, whereas running $\tilde{R}\ll R$ iterations of ABC-DC over $M>1$  processors is feasible (for simplicity, assume $K$ a multiple of $M$). Here the task of computing the cloned likelihood would be performed by distributing $K/M$ model simulations to each of the $M$ processors.

In our examples timing is obtained on simulations running on a i7-4790 CPU 3.60 GHz PC desktop.

\subsection{$g$-and-$k$ distribution}\label{sec:g-and-k}

An interesting case study is given by $g$-and-$k$ distributions, first analysed via ABC methods in \cite{allingham2009bayesian}. This is a flexibly shaped distribution that is used to model non-standard data
through a small number of parameters. It is defined by its inverse distribution
function, but has no closed form density. The quantile function (inverse distribution function) is given by 
\begin{equation}
F^{-1}(x;A,B,c,g,k)= A+B\biggl[1+c\frac{1-\exp(-g\cdot r(x))}{1+\exp(-g\cdot r(x))}\biggr](1+r^2(x))^kr(x)
\label{eq:g-k-inverse}
\end{equation}
where $r(x)$ is the $x$th standard normal quantile, $A$ and $B$ are location and scale parameters and $g$ and $k$ are related to skewness and kurtosis. We assume $\theta=(A,B,g,k)$ as parameter of interest, given that we keep $c$ fixed to $c=0.8$ \citep{rayner2002numerical}. Parameters restrictions are $B>0$ and $k>-0.5$.
An evaluation of \eqref{eq:g-k-inverse} returns a draw ($x$th quantile) from the $g$-and-$k$ distribution or, in other words, the $i$th sample $r_i:=r_i(x)\sim N(0,1)$ produces a draw $z_i:=F^{-1}(\cdot;A,B,c,g,k)$ from the $g$-and-$k$ distribution.  Notice in this case there is no hidden/latent process, hence all simulated values $z_i$ are independent draws from said distribution.

\begin{figure}
\centering
\includegraphics[width=17cm,height=9cm]{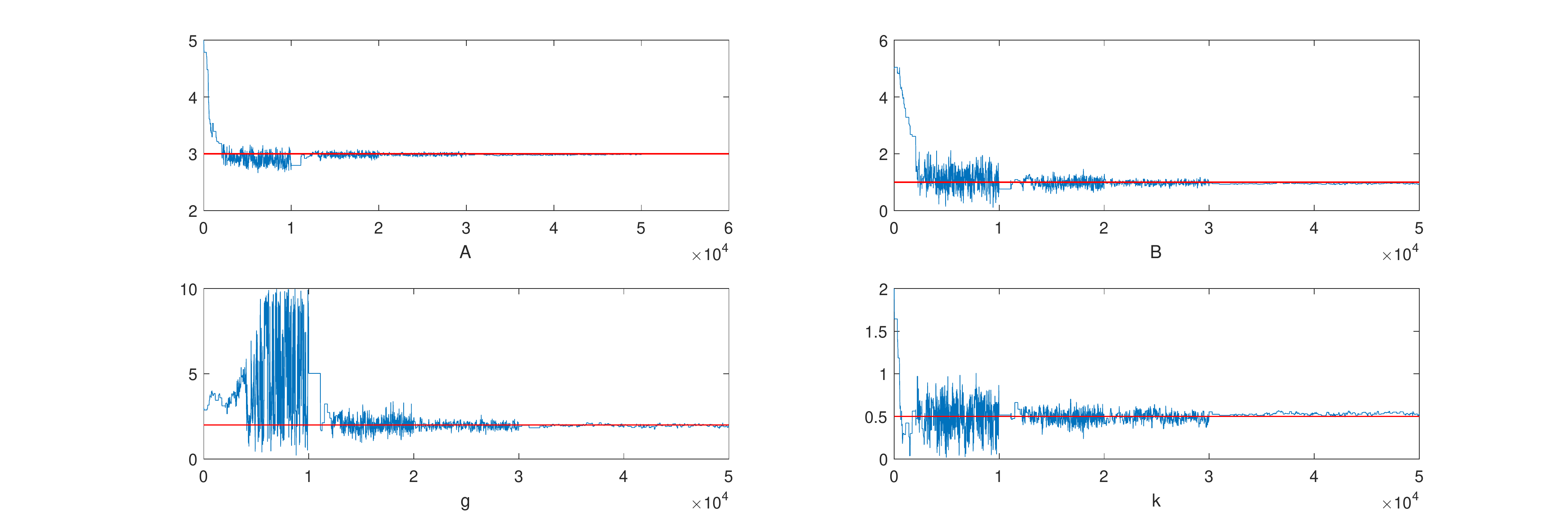}
\caption{\footnotesize{g-k distribution: trace plots for the ABC-MCMC run. A major variability reduction occurs at iteration 10,000 when reducing $\delta$ from 0.3 to 0.1. Horizontal lines give the true parameter values.} }
\label{fig:g-k_abcmcmc_trace}
\end{figure}

We follow the simulation setup for data $(y_1,...,y_n)$ of size $n=10^4$ generated as in \cite{allingham2009bayesian} (see also \citealp{fearnhead-prangle(2011)}) with $\theta=(3,1,2,0.5)$. They set uniform priors $U(0,10)$ on each parameter then used a standard ABC-MCMC algorithm (\citealp{marjoram2003markov}) to estimate parameters, with summaries $S(y)=(y_{(1)},...,y_{(n)})$ (the sequence of ordered data) and $J(y,z)=(\sum_{i=1}^n[S_i(z)-S_i(y)]^2)^{1/2}$ with $S_i$ the $i$th element of $S$ and $z=(z_1,...,z_n)$ a vector of samples from the $g$-and-$k$ distribution. They obtain good inference for all parameters but $g$ which is essentially unidentified. Actually it is extremely simple to obtain accurate inference for all parameters by reducing the dimensionality of the problem using a smaller set of summaries. We set $S(y)=(P_{20},P_{40},P_{60},P_{80},\mathrm{skew}(y))$, that is the 20-40-60-80th percentiles of the data and the sample skewness. As comparison function $J_{\delta}$ we consider a different criterion, namely a Gaussian kernel
as in \eqref{eq:gauss-kernel}. With such setup, we first run a standard ABC-MCMC without data-cloning and let $\delta$ decrease with schedule $\delta\in\{5,3,1\}$. Results were not encouraging, and that's because of using a matrix of weights $\Omega$ with unit diagonal thus giving the same weight to each of the five summary statistics. Then we formed a new matrix of weights from the output of such preliminary (pilot) run, as described in section \ref{sec:abc-basics}, to obtain $[\omega_1,...,\omega_5]=[0.22,    0.19,    0.53,    2.96,    1.90]$.
With the new $\Omega$ we run ABC-MCMC once more, this time with
$\delta\in\{0.3, 0.1, 0.05, 0.015\}$ where the largest value of $\delta$ was used for the first 10,000 iterations, then decreased every 10,000 iterations and the smallest value was used for the last 20,000 iterations of overall $R=50,000$ ABC-MCMC iterations, starting from parameter values (5,5,3,2). The 50,000 iterations were completed in 103 seconds. Trace plots are in Figure \ref{fig:g-k_abcmcmc_trace}. 
At the smallest $\delta=0.015$ we obtain an acceptance rate of 1-2\% and corresponding posterior means and 95\% posterior intervals: $A=2.98$ (2.97,2.99), $B=0.95$ (0.92,0.98), $g=1.95$ (1.82,2.07), $k=0.53$ (0.50,0.56). 

\begin{figure}
\centering 
\includegraphics[width=17cm,height=9cm]{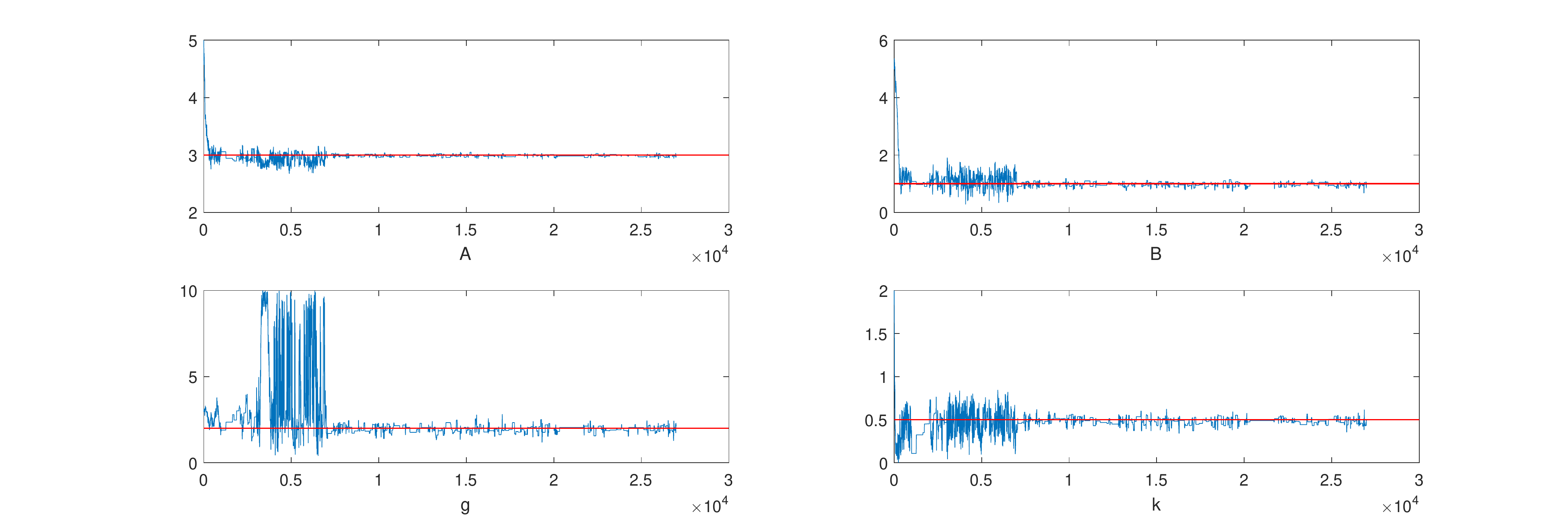}
\caption{\footnotesize{g-k distribution: trace plots for the ABC-DC run. Here $\delta=0.3$ constantly and the variability reduction at iteration 7,000 is only due to $K$ increasing from $K=1$ to $K=15$. Horizontal lines give the true parameter values.}}
\label{fig:g-k_abcdc_trace}
\end{figure}

\begin{figure}
\centering
\includegraphics[scale=0.4]{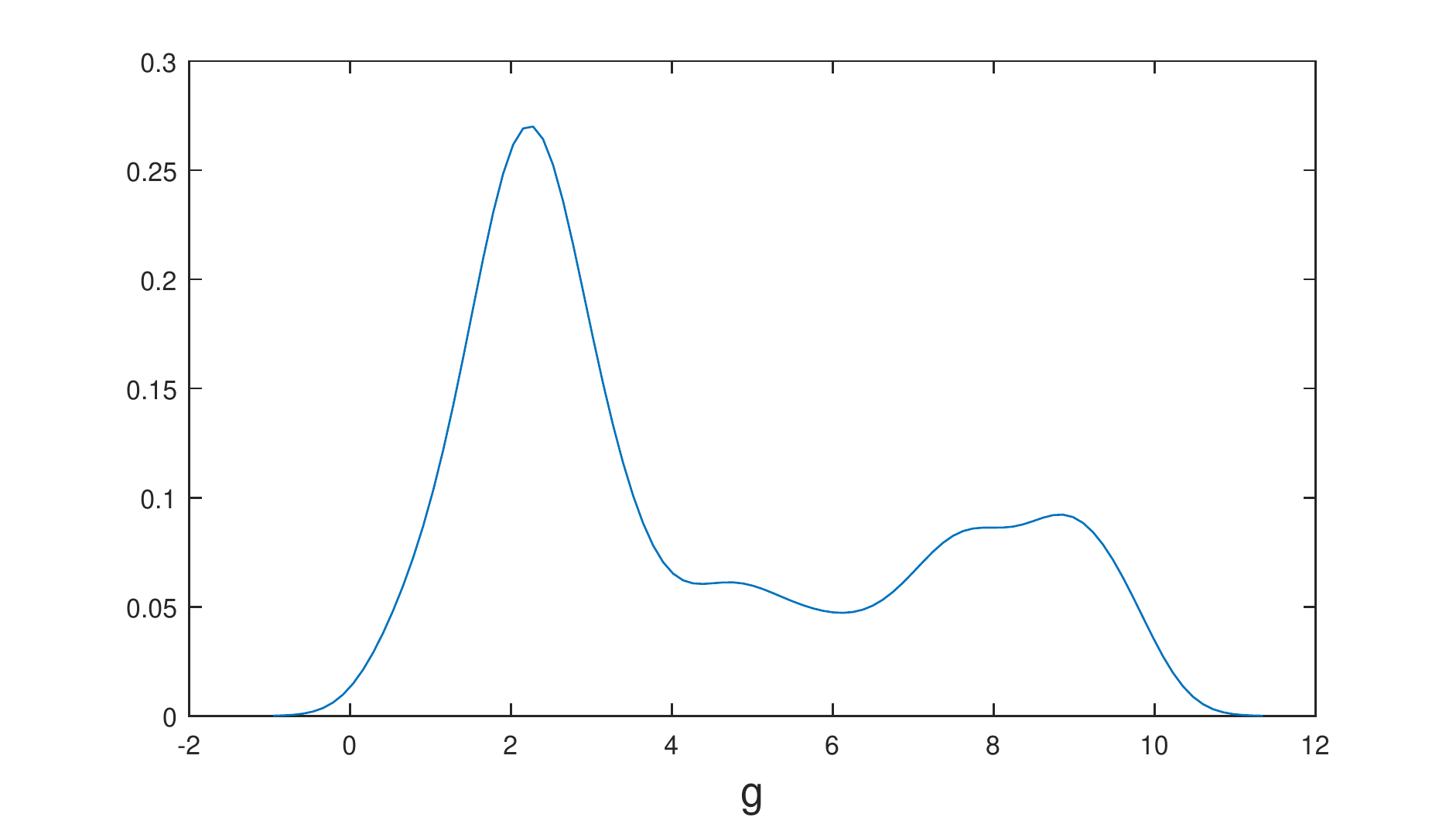}
\caption{\footnotesize{g-k distribution: marginal posterior for $g$ when $K=1$ and $\delta=0.3$. At this point the current main mode $\tilde{\theta}$ is automatically identified by ABC-DC and the algorithm will use it to propose samples under a larger value of $K$.}}
\label{fig:g-k_density_gparameter}
\end{figure}

Considering a pilot run was doubly useful, because we can now use the determined $\Omega$ into ABC-DC. We start the algorithm at the same starting parameter values used in ABC-MCMC and keep the threshold fixed to a large value, that is $\delta=0.3$ and in this case we do not let it decrease.  Notice that $\delta=0.3$ is the largest value used in the previous ABC-MCMC experiment. The first 7,000 iterations are run with $K=1$, useful to identify a temporary main mode $\tilde{\theta}$ (see Figure \ref{fig:g-k_density_gparameter}), then we enlarge it to $K=15$ and during the next 20,000 iterations observe an acceptance rate of 1-2\%. Therefore we ran in total $R=27,000$ iterations which are completed in 402 seconds.  Trace plots are in Figure \ref{fig:g-k_abcdc_trace}. 
Using $(\delta,K)=(0.3,15)$ we obtain the following asymptotic means and standard errors from the large samples arguments outlined in section \ref{sec:dynamic-abcdc}: $\hat{A}=2.99$ (0.07), $\hat{B}=0.98$ (0.26), $\hat{g}=1.97$ (0.77), $\hat{k}=0.48$ (0.14). As previously remarked, in the present work we focus on point estimation and in this section confidence intervals based on asymptotics are reported only to show that these are likely to be overestimated when using ABC-DC, see the discussion below where we also consider a bootstrap approach.
We considered the setup above for ABC-DC to perform a fair comparison with ABC-MCMC, namely an $R$ long enough to return 20,000 draws (when $K=15$) and a $K$ large enough to return the same acceptance rate as in ABC-MCMC. Of course the setting is time consuming, however we can show how to obtain the same results using a quicker ABC-DC. We run 7,000 iterations with $(\delta,K)=(0.3,1)$ then increase the number of clones to $K=5$ for further 5,000 iterations; overall time is 48 seconds and the acceptance rate is 10-12\%. We obtain $\hat{A}=2.98$ (0.06), $\hat{B}=0.97$ (0.23), $\hat{g}=1.99$ (0.74), $\hat{k}=0.49$ (0.16) essentially the same results obtained under the more expensive setup with computations an order of magnitude faster. For both ABC-DC simulations it is interesting to appreciate how the variability of the chain for parameter $g$ (when $K=1$), and the automatic identification of its maximum, allowed for a dramatic improvement in the mode identification for a larger $K$. See Figure \ref{fig:g-k_density_gparameter} for the approximate marginal distribution of $g$ based on $K=1$: ABC-DC automatically identifies the highest mode at about $g=2$ and focuses on such mode for an increasing $K$.

We conclude that ABC-DC returns very good point estimates, however standard errors are likely to be overestimated. To verify the latter claim we run a parametric bootstrap procedure of size 100: that is we produce 100 independent datasets using the parameters obtained under the more computationally conservative approach ($\hat{A}=2.98$, $\hat{B}=0.97$, $\hat{g}=1.99$, $k=0.49$ when $K=5$ and $\delta=0.3$) then re-estimate parameters on each simulated dataset via ABC-DC. For each simulation we used, again, 7,000 iterations with $(\delta,K)=(0.3,1)$ followed by 5,000 iterations with $(\delta,K)=(0.3,5)$. Bootstrap results are in Table \ref{tab:g-and-k-bootstrap}, showing reasonable variation, whereas previously standard errors built on asymptotic arguments are overestimated. Finally, we perform a parametric bootstrap procedure employing an exact maximum likelihood estimation obtained by numerical optimization \citep{rayner2002numerical}, using the R package in \url{https://github.com/dennisprangle/gk}. Results are in Table \ref{tab:g-and-k-bootstrap}, and we can once more appreciate the good approximation provided by ABC-DC. Thanks to this further simulation, we can definitely confirm that trying to apply the asymptotic considerations from section \ref{sec:dynamic-abcdc} to compute standard errors (instead of, say, bootstrap procedures) results in inflated confidence intervals.
\begin{table} 
\centering
\caption {g-k distribution: means over 100 parametric bootstrap replications and 2.5-97.5\% empirical percentiles using an exact MLE procedure and ABC-DC. For each replication using ABC-DC we considered $(\delta,K)=(0.3,5)$.}
\begin{tabular}{cccc}
\hline
\hline
 & True values & MLE &  ABC-DC\\ 
\hline 
$A$ & 2.98 & 2.99 [2.97,3.01] & 2.98 [2.95,3.01]\\
$B$ & 0.97 & 0.98 [0.95,1.02] & 0.98 [0.91,1.04]\\
$g$ & 1.99 & 1.97 [1.92,2.02] & 2.15 [1.89,2.70]\\
$k$ & 0.49 & 0.49 [0.47,0.51] & 0.48 [0.41,0.58]\\
\hline
\end{tabular}
\label{tab:g-and-k-bootstrap}
\end{table}

\subsection{Stochastic Gompertz model}\label{sec:gompertz}

Here we consider a state-space model with stochastic Gompertz dynamics. \cite{donnet2010bayesian} and \cite{ditlevsen2013introduction} used a hierarchical (mixed-effects) version of this model to study chicken growth. We do not use a mixed-effects model so our results cannot be directly compared to the cited references. We have the stochastic differential equation (SDE)
\begin{equation}
dX_t = BCe^{-Ct}X_tdt + \sigma X_tdW_t,\qquad X_0=Ae^{-B}
\label{eq:gompertz-state}
\end{equation}
with $A$, $B$, $C$ and $\sigma$ unknown positive constants. It is easy to prove by It\^{o}'s formula on the transformed process $Z_t=\log(X_t)$ that \eqref{eq:gompertz-state} has explicit solution $X_t=Ae^{-(Be^{-Ct})-\frac{1}{2}\sigma^2t+\sigma W_t}$ and $X_0=Ae^{-B}$. Same as in \cite{donnet2010bayesian} and \cite{ditlevsen2013introduction} we consider data on a logarithmic scale according to 
\[
(y_0,y_1,...,y_n) = (\log(A)-B,\log(X_1),...,\log(X_n))+\epsilon 
\]
where $\epsilon\sim N_{n+1}(0,\sigma^2_{\epsilon}I_{n+1})$ has $(n+1)$-dimensional multivariate Gaussian distribution (here $I_{n+1}$ is the identity matrix). For simplicity we assume $X_0$ known hence an estimate of either $A$ or $B$ can be determined from an estimate of the other parameter, e.g. $B=\log(A/X_0)$. In the following we choose to determine $B$. We assume $\sigma_{\epsilon}$ known as this is a difficult parameter to identify without access to repeated measurements. Therefore unknowns are $(A,C,\sigma)$ and, in order to preserve positivity, in practice we conduct inference for $\theta=(\log A,\log C,\log\sigma)$. We set the following priors $\log A\sim U(1,15)$, $\log C \sim U(0.5,4)$, $\sigma \sim LN(0.1,0.2)$. Here $LN(a,b)$ denotes the log-Normal distribution with parameters $(a,b)$ ($a$ and $b$ being the mean and standard deviation respectively of the associated Normal distribution). 

We simulate $n+1=51$ data points at equispaced observational times $(t_0,...,t_n)=(0,1,...,50)$ with parameters $(\log A,\log B,\log C,\log\sigma,\log\sigma_{\epsilon})=(8.01,1.609,2.639,0,-1.609)$. In practice we normalize times to be in [0,1] for numerical stability.
We wish to consider summary statistics to ease inference via ABC, however the determination of summaries for dynamic models is way less intuitive than for static models. We employ the regression approach suggested in \cite{fearnhead-prangle(2011)} to determine a set of three summaries $S(\cdot)=(S_1(\cdot),S_2(\cdot),S_3(\cdot))$. Essentially their ``semi-automatic ABC'' is such that the vector $S(\cdot)$ has the same dimension as $\theta$, that is each element of $S(\cdot)$ is supposed to be informative for a given component of $\theta$. We do not illustrate their method here, but the reader can refer to \cite{picchini-2013} for an exposition targeting SDE models and to our MATLAB package \citep{abc-sde} implementing the Fearnhead-Prangle method for the determination of $S(\cdot)$ (but not implementing data-cloning).  

\begin{figure}
\centering
\includegraphics[scale=0.7]{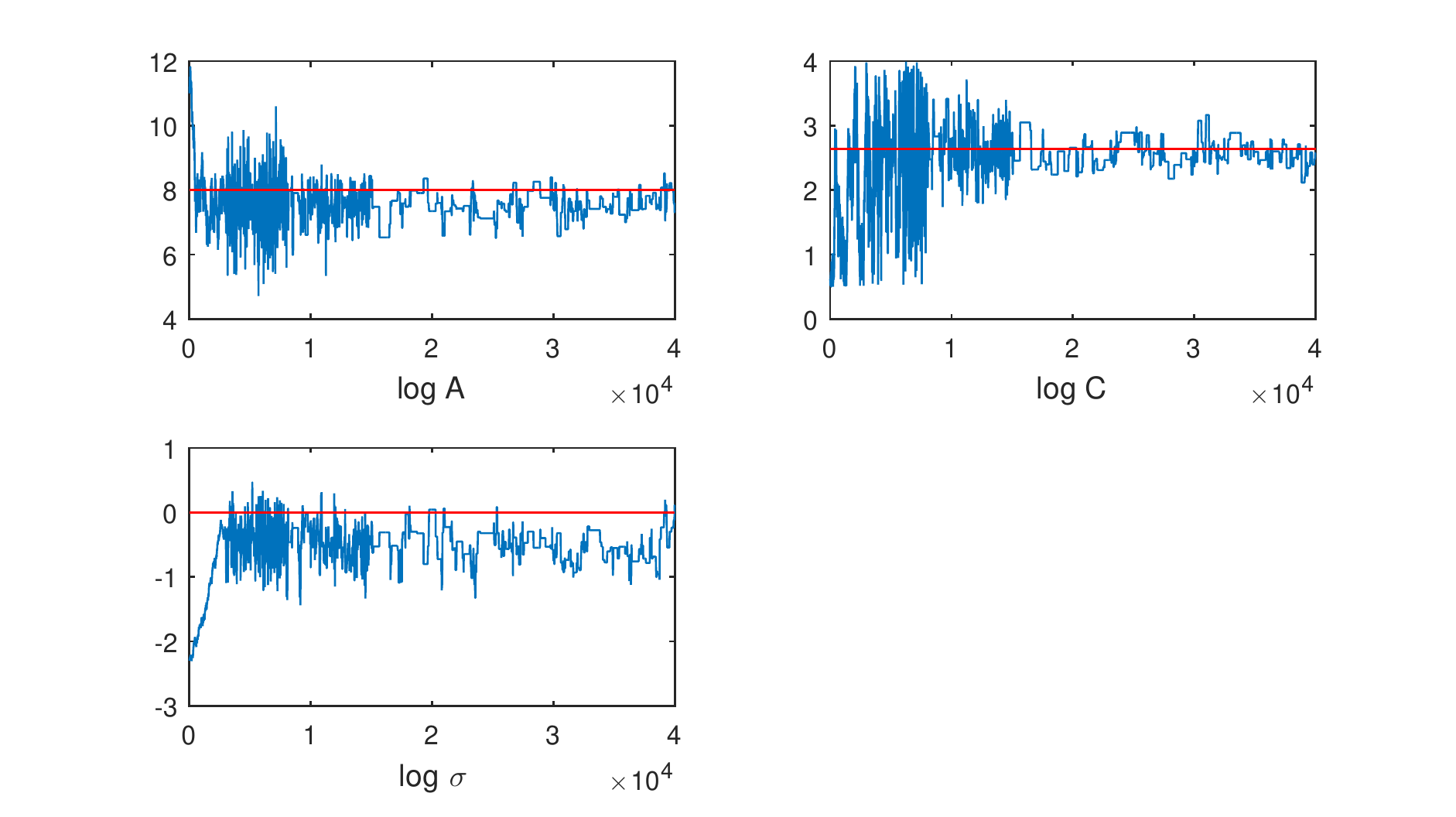} 
\caption{\footnotesize{Gompertz model: traceplots for $\log A$, $\log C$ and $\log \sigma$ when using ABC-MCMC. Horizontal lines are parameter true values.}}
\label{fig:gompertz_abcmcmc_traces}
\end{figure}

We first illustrate the results from an ABC-MCMC without data-cloning: we start at initial parameter values $(\log A_0,\log C_0,\log\sigma_0)=(11,0.6,-2.3)$ and first run a pilot study to determine weights $(\omega_1,\omega_2,\omega_3)$ weighting the three automatically obtained summaries $(S_1,S_2,S_3)$. Then we run ABC-MCMC once more using these weights and produce a total of $R=40,000$ draws. The first 8000 iterations use $\delta=20$, then we decrease it to $\delta=5$ for 7000 iterations and finally to $\delta=1.5$ for the last 25000 iterations with a 1\% acceptance rate at the smallest threshold. Automatic summaries construction together with the 40,000 ABC-MCMC iterations required about 55 seconds. Trace plots are in Figure \ref{fig:gompertz_abcmcmc_traces}. Posterior means are $\log\hat{A}=7.52$, $\log\hat{C}=2.58$, $\log\hat{\sigma}=-0.52$. As we see in a moment the identification of $\sigma$ is difficult because we can't really make use of an informative summary statistic.

\begin{figure}
\centering
\includegraphics[scale=0.7]{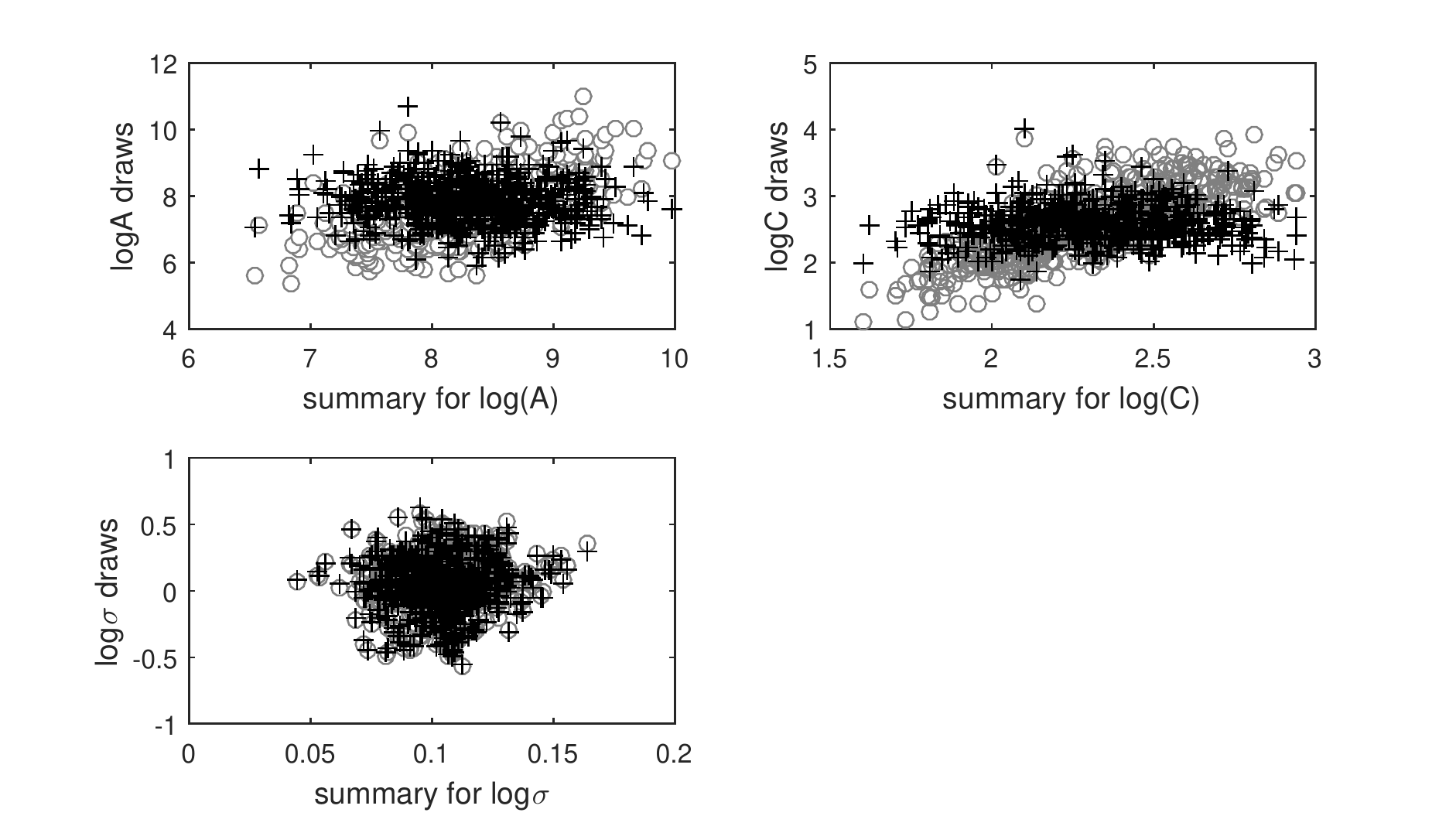} 
\caption{\footnotesize{Gompertz model: ABC-DC with $K=1$. Abscissas have values of simulated summary statistics and ordinates have the corresponding simulated parameters, labelled with gray circles (o). Black plusses (+) denote corresponding regularized parameters.}}
\label{fig:gompertz-regularization-top}
\end{figure}

\begin{figure}
\centering
\includegraphics[scale=0.7]{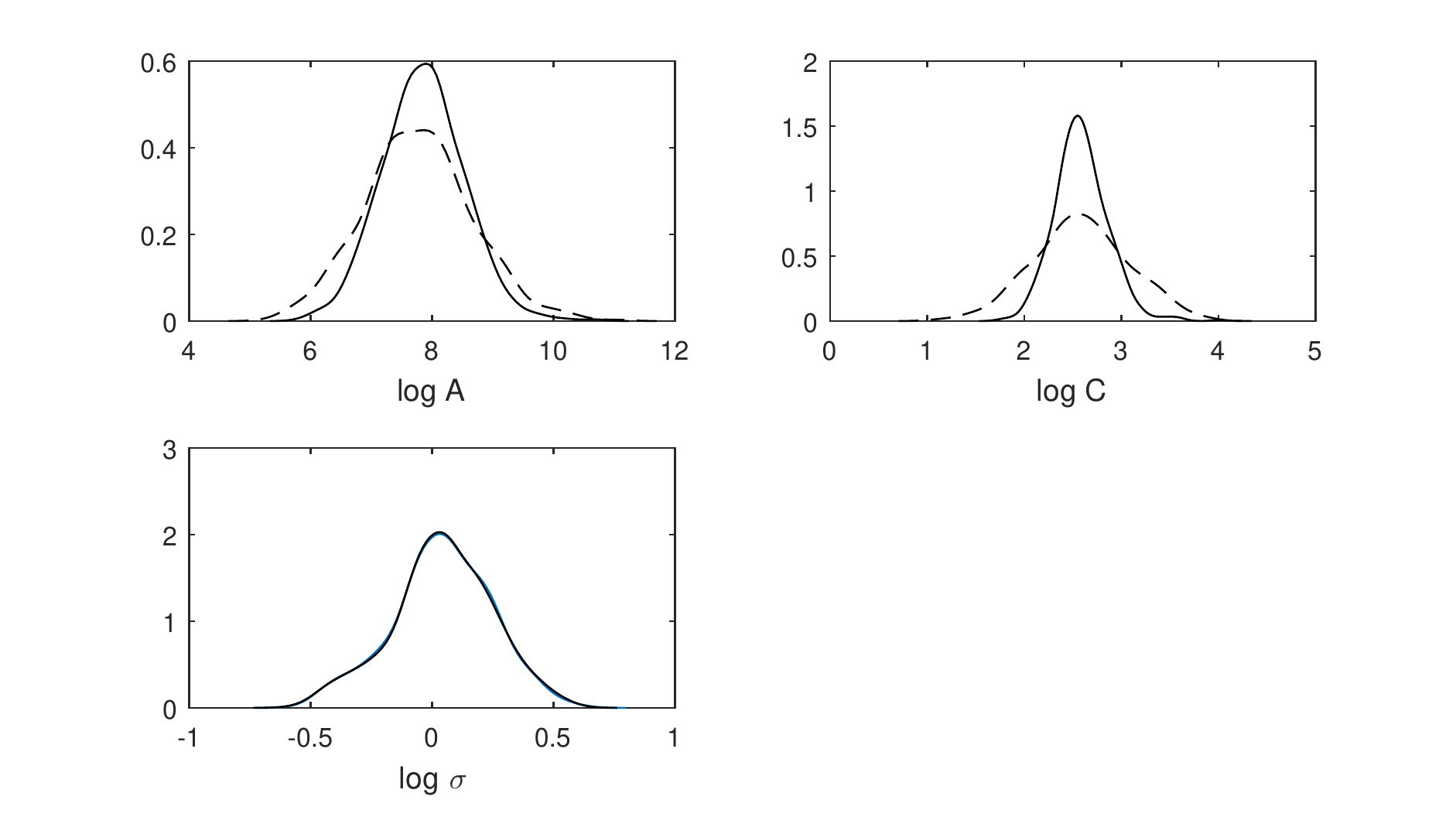} 
\caption{\footnotesize{Gompertz model: ABC-DC with $K=1$. Kernel smoothed marginals for regression adjusted parameters (solid lines) and from the original non-adjusted draws (dashed lines). For $\log\sigma$ the two marginals superimpose.}}
\label{fig:gompertz-regularization-densities}
\end{figure}

\begin{figure}
\centering
\includegraphics[scale=0.7]{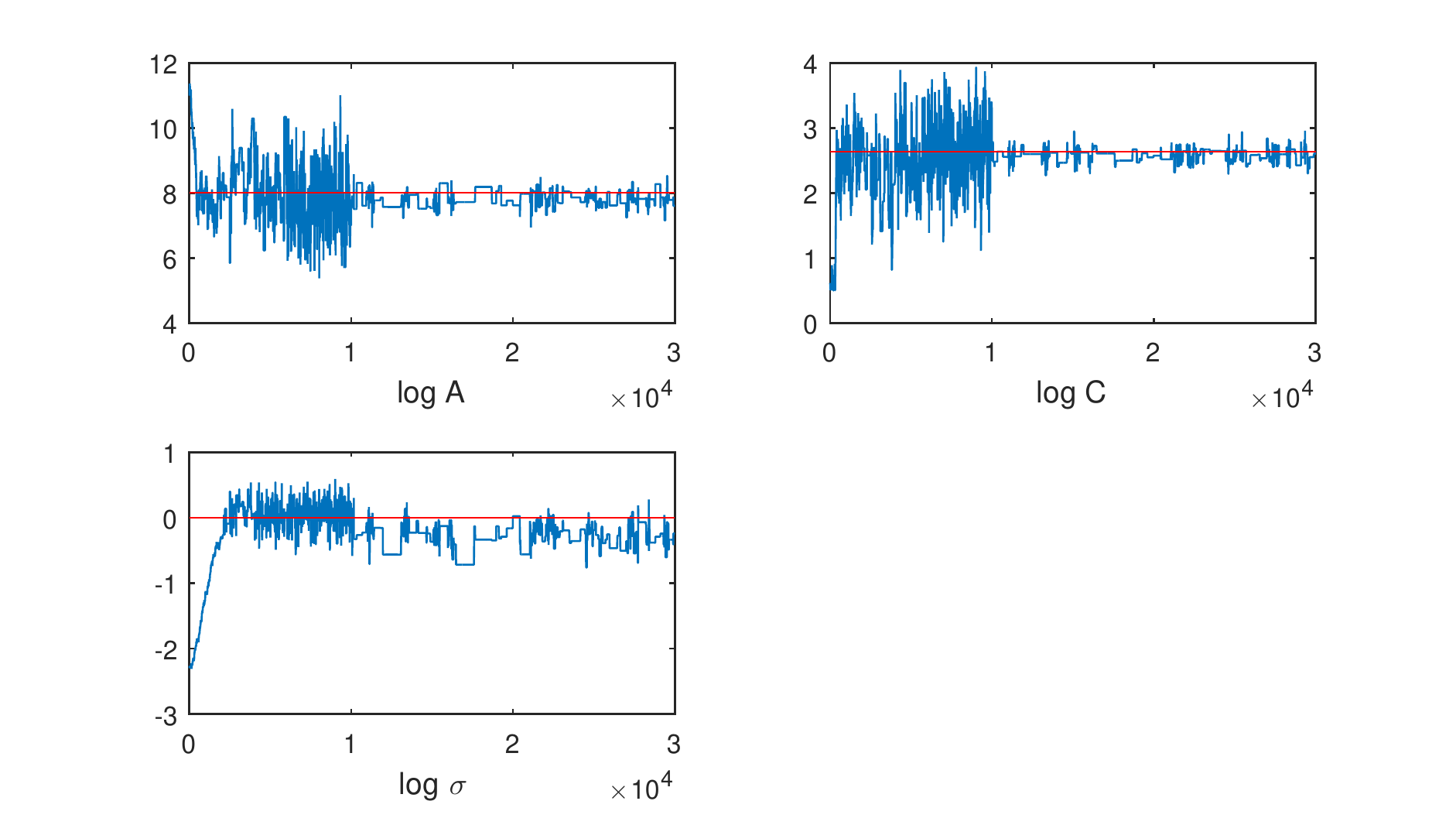} 
\caption{\footnotesize{Gompertz model: traceplots for $\log A$, $\log C$ and $\log \sigma$ when using ABC-DC. Horizontal lines are parameter true values.}}
\label{fig:gompertz_abcdc_traces}
\end{figure}

We now consider ABC-DC where $K=1$ is used for the first 10,000 iterations and we keep $\delta=18$ constant for the entire simulation, i.e. this $\delta$ is about thirteen times larger than the smallest $\delta$ used in ABC-MCMC. At the end of the 10,000 iterations we apply the regression adjustment described in section \ref{sec:regression-adjustment}. This is trivially implemented and has negligible impact on the overall computational budget, amounting to solve a system of linear equations using draws already sampled at previous iterations, i.e. it is an almost instantaneous operation. In Figure \ref{fig:gompertz-regularization-top} circles denote pairs $(\theta_i,S_i)$ where the $\theta_i$ are draws generated with $K=1$ (after burnin) and $S_i$ the corresponding three-dimensional simulated summary statistic, one for each dimension of $\theta_i$. Plusses denote the pairs $(\theta_i^*,S_i)$, i.e. regression-adjusted parameters. We deduce that regularization enables an improved identification of $A$ and $C$, which is of great help given the large value of $\delta$ we are using. However the summary statistic used for $\sigma$ seems totally uninformative for the said parameter and in fact the regularization has no effect, see also Figure \ref{fig:gompertz-regularization-densities}. As explained in section \ref{sec:regression-adjustment} we can use the sample covariance from the adjusted draws to create a more effective independence sampler when $K>1$. We employ this strategy here, and furthermore we center the independence sampler to the  mean of the adjusted draws $\theta^*_i$ and execute 20,000 further iterations using $K=11$ clones. Trace plots are in Figure \ref{fig:gompertz_abcdc_traces}. Sample means computed on the last 20,000 draws returns point estimates $\log\hat{A}=7.88$, $\log\hat{C}=2.71$, $\log\hat{\sigma}=-0.288$ which are closer to the true parameter values than those obtained via ABC-MCMC.

\begin{figure}
\centering
\includegraphics[scale=0.7]{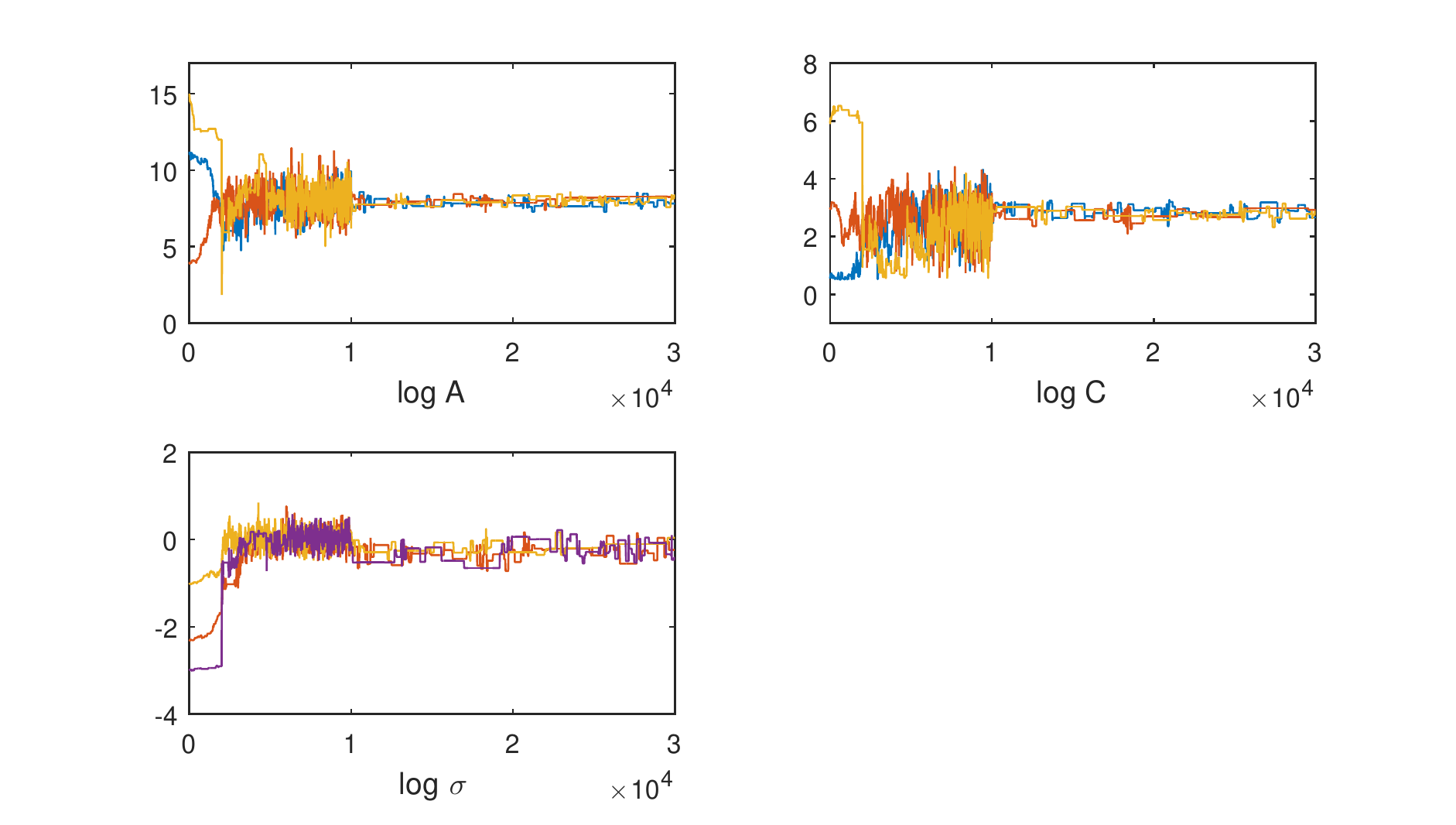} 
\caption{\footnotesize{Gompertz model: three independent chains of ABC-DC starting at different values.}}
\label{fig:gompertz-threestartingvalues}
\end{figure}

\begin{figure}
\centering
\includegraphics[scale=0.4]{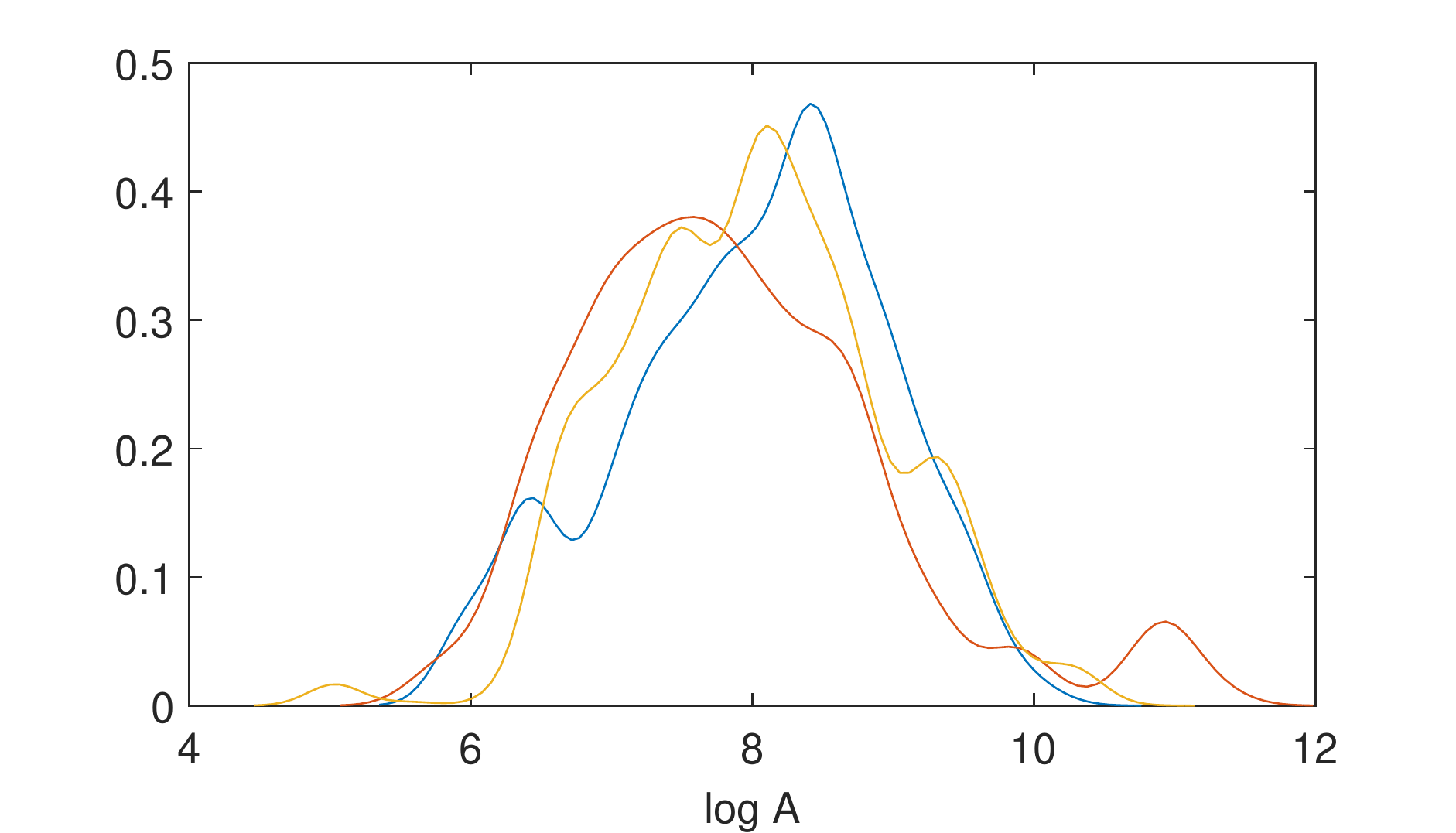} 
\caption{\footnotesize{Gompertz model: marginal posteriors without regression adjustment for each of three chains for $\log A$ resulting from iterations 7,000-10,000 in Figure \ref{fig:gompertz-threestartingvalues}, that is from $K=1$ and $\delta=18$. See main text for comments.}}
\label{fig:gompertz-logA-1clone-density}
\end{figure}

With $K=11$ we obtain a 1\% acceptance rate, for comparison with inference via ABC-MCMC. The entire algorithm (including the calculation of summary statistics and the regression adjustment) required about 48 seconds. This was possible thanks to a carefully vectorized MATLAB code so that simulating multiple instances of our model does not have any significant impact on the computational performance. Notice that without adjustment we are not able to jump from $K=1$ straight to $K=11$, as this would result in a extremely low acceptance rate. Finally, in order to show the robustness of the method, we perturb the parameters starting values and produce three different chains given in Figure \ref{fig:gompertz-threestartingvalues}, all converging to the same values despite the large ABC tolerance and very different starting parameters. For example, in Figure \ref{fig:gompertz-logA-1clone-density} we show the marginal posteriors for $\log A$ for the three chains resulting from $K=1$ and before applying regression adjustment: as we can see the main modes are rather different. Despite this, for increasing $K$ the three ABC-DC chains converge to about the same value, as we have just discussed.

Our main comment is that we obtain a reasonable point estimate using ABC-DC without having to reduce $\delta$ too much, which should be particularly relevant for even more complex modelling scenarios, as commented at the beginning of section \ref{sec:simulations}.

\subsection{Two dimensional correlated Geometric Brownian motion }\label{sec:2GBM}

In this section we apply ABC-DC to estimate the parameters of a two-dimensional SDE model having a likelihood function analytically available. It is therefore possible to compare results based on our methodology with exact maximum likelihood estimation. The model considered is a two-dimensional geometric Brownian motion or Black-Scholes model, the standard asset price model in finance.
The process $\left\{X_{t},Y_{t}\right\}_{t\geq0}$ is solution of
the system of SDEs
\begin{align}
\begin{cases}
 dX_{t}&=\mu_{1}X_{t}dt+\sigma_{1}X_{t}dW_{t}^1 \\
 dY_{t}&=\mu_{2}Y_{t}dt+\sigma_{2}Y_{t}dW_{t}^2
\label{eq:GBM}
\end{cases}
\end{align}
with $\sigma_{1},\sigma_{2}>0$, $\{W^{1}_t\}$ and $\{W^{2}_t\}$ correlated Wiener processes such that $dW_{t}^{1}\cdot dW_{t}^{2}=\rho \cdot dt$ with a correlation coefficient $|\rho|<1$.
We can explicitly write the two components of the exact solution for 
$t\in[0,\infty)$ in terms of two independent Wiener processes $\{B^{1}_t\}$
and $\{B^{2}_t\}$ as 
\begin{align*}
\ X_{t}&=X_{0}\exp\left\{ \left(\mu_{1}-\frac{1}{2}\sigma_{1}^{2}\right)t+\sigma_{1}B_{t}^{1}\right\} \\
\ Y_{t}&=Y_{0}\exp\left\{ \left(\mu_{2}-\frac{1}{2}\sigma_{2}^{2}\right)t+\sigma_{2}\left(\rho B_{t}^{1}+\sqrt{1-\rho^{2}}B_{t}^{2}\right)\right\} 
\end{align*}
where $ (X_{0},Y_{0}) $ is the deterministic starting state of the process.
By application of the multi-dimensional It{\^o} formula we can derive an explicit expression for the conditional transition densities. It follows that for $s\in\left[0,t\right),\, t\in[0,\infty),$ 
the transition density is 

\[
p(s,x_{s},y_{s};t,x_{t},y_{t})=\frac{1}{2\pi(t-s)\sqrt{1-\rho^{2}}\sigma_{1}\sigma_{2}x_{t}y_{t}}\cdot
\]
\[
\exp\biggl(-\frac{(\ln(x_{t})-\ln(x_{s})-(\mu_{1}-\frac{1}{2}\sigma_{1}^{2})(t-s))^{2}}{2\sigma_{1}^{2}(t-s)(1-\rho^{2})}
\]

\[
-\frac{(\ln(y_{t})-\ln(y_{s})-(\mu_{2}-\frac{1}{2}\sigma_{2}^{2})(t-s))^{2}}{2\sigma_{2}^{2}(t-s)(1-\rho^{2})}+
\]

\[
\frac{(\ln(x_{t})-\ln(x_{s})-(\mu_{1}-\frac{1}{2}\sigma_{1}^{2})(t-s))(\ln(y_{t})-\ln(y_{s})-(\mu_{2}-\frac{1}{2}\sigma_{2}^{2})(t-s))\rho}{\sigma_{1}\sigma_{2}(t-s)(1-\rho^{2})}\biggr).
\]
We assume observations generated from model \eqref{eq:GBM}, hence due to the Markovian property of the model solution it is possible to write the exact likelihood function for a discrete sample from the process as proportional to the product of its transition densities. MLEs are therefore obtained by maximizing numerically the resulting likelihood function, see Table \ref{tab:GBM-abcdc}. 

We wish to conduct inference for the set of parameters $\theta = \left(\mu_1, \ln(\sigma_1), \mu_2, \ln(\sigma_2), \rho \right)$ using $500$ equispaced observations taken on the time interval  $\left[0,1\right]$. The starting state of the model is set to $({X}_{0}, {Y}_{0})=\left(1,2\right)$ and the true values of the parameters are $\mu_{1}=1.7$, $\ln\sigma_{1}=-0.8$, $\mu_{2}=1.3,$ $\ln\sigma_{2}=-1.2$ , $\rho=0.3$.
The dimension of the problem creates difficulties in comparing data and pseudo-data, since to preserve a reasonable acceptance rate we would need to fix the threshold $\delta$ to high values. However for this example we are able to identify sufficient statistics, resulting in much higher acceptance rates compared to using the entire dataset. Sufficiency implies that comparing summaries of data and pseudo-data is equivalent to comparing actual and simulated data. 
We denote with $\left(M_{1},V_{1},M_{2},V_{2},R_{1},R_{2}\right)$ the vector
of sufficient statistics based on discrete observations $\{Z_i\}_{i=0,..,n}=\{X_{i},Y_{i}\}_{i=0,..,n}$. The statistics are
\begin{align*}
\  M_{1}&=\sum_{i=1}^{n}(\ln X_{i}-\ln X_{i-1})=(\ln X_{n}-\ln X_{0}) ,
\  V_{1}=\sum_{i=1}^{n}(\ln X_{i}-\ln X_{i-1})^{2}  \\
\ M_{2}&=\sum_{i=1}^{n}(\ln Y_{i}-\ln Y_{i-1})=(\ln Y_{n}-\ln Y_{0})  ,
\ V_{2}=\sum_{i=1}^{n}(\ln Y_{i}-\ln Y_{i-1})^{2}  \\
\ R_{1}&=\sum_{i=1}^{n}(\ln X_{i}-\ln X_{i-1})(\ln Y_{i}-\ln Y_{i-1})  ,
\ R_{2}=\sum_{i=1}^{n}\ln(X_{i}Y_{i}).  
\end{align*}

Since these statistics vary on different scales, we weight them using the estimated standard deviation $\hat{\sigma}_{j}$ of each statistic, obtained using a pilot run of ABC-MCMC considering 20,000 iterations and using thresholds $\delta=\left(1,0.9,0.8\right)$, updated at iterations 7,000 and 14,000. We denote with ${S}\left(z\right)=\left(S_{j}(z)\right)_{j=1,..6}$ the vector of statistics for a dataset $z$ and with $S(z^*)$ the corresponding quantity for a simulated $z^*$. By considering for $J_{\delta}\left({S}(z),{S}(z^{*})\right)$ a Gaussian kernel
$
\prod_{k=1}^{K}\exp\left(-u/(2\delta^{2})\right)/\delta
$, we weight summary statistics by writing $u={D^{T}}{\Sigma}{D}$ where
${D^{T}}=\left({S}(z^{*(k)})-{S}(z)\right)^{T}$
and diagonal matrix ${\Sigma}=\mathrm{diag}\left(\omega_{1},...,\omega_{6}\right)$
with $\omega_{j}=1/\hat{\sigma}^2_j$. 

We conduct inference on the logarithms of the two volatility parameters $\sigma_{1}$ and $\sigma_{2}$, since they are meant to be strictly positive. Moreover, the correlation parameter $\rho$ should be in $(-1,1)$, therefore we truncate its prior on this interval. We set priors ${N}\left(1.5,0.5^2\right)$ on the drift parameters $\mu_{1}$ and $\mu_{2}$, whereas the priors on $\ln\sigma_{1}$ and $\ln\sigma_{2}$ are chosen to be ${N}\left(-1,0.5^2\right)$, and as prior on
$\rho$ we set a truncated Gaussian $N_{(-1,1)}\left(0.5,0.3^2\right)$ with subscript denoting the truncation at the specified interval. The starting values for the parameters are set to $\theta_0=\left(1.5,-1,1.5,-1,0.1\right)$. 
We compare the estimates obtained using ABC-MCMC and ABC-DC on a total of 100,000 iterations. In both cases, the covariance for the adaptive Metropolis algorithm is updated every 1,000 iterations. 

\begin{figure}
\centering
\includegraphics[scale=0.58]{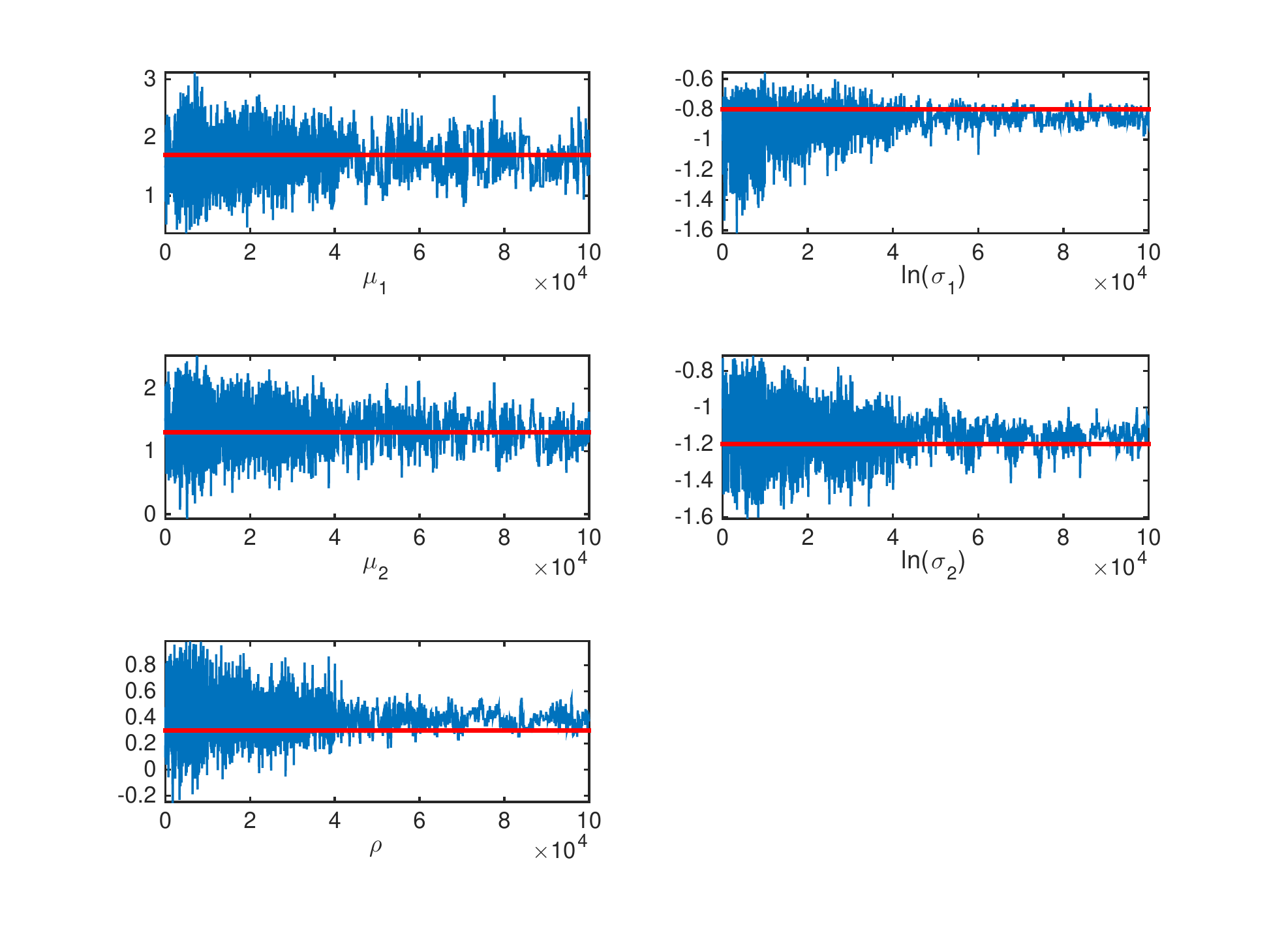}
\vspace{-1cm}
\caption{Geometric Brownian motion: (top) traceplots for $\mu_1$, $\log \sigma_1$; (middle) $\mu_2$, $\log\sigma_2$ and (bottom) $\rho$ when using ABC-MCMC and updating progressively the threshold $\delta$ to the values  $\left(0.8,0.5,0.4,0.3,0.2,0.15\right)$. Horizontal lines are the true parameters.}
\label{fig:GBM_ABC}
\end{figure}

With ABC-MCMC the threshold  $\delta$ is updated every 10,000 iterations within the values  $\left(0.8,0.5,0.4,0.3,0.2\right)$, then after further 20,000 iterations  is decreased to $\delta=0.2$, and finally from iteration 60,000 to the end of the simulation the threshold is fixed at $\delta = 0.15$. During this last stage the acceptance ratio varies between 1 and 4\%. After a total running time of 198 seconds,  we obtain the estimated posterior means $ \hat{\theta} = (1.6426 , -0.8495 ,1.2979,  -1.1599 ,   0.3860)$. Traceplots are given in Figure \ref{fig:GBM_ABC}.

\begin{figure}
\centering
\includegraphics[scale=0.58]{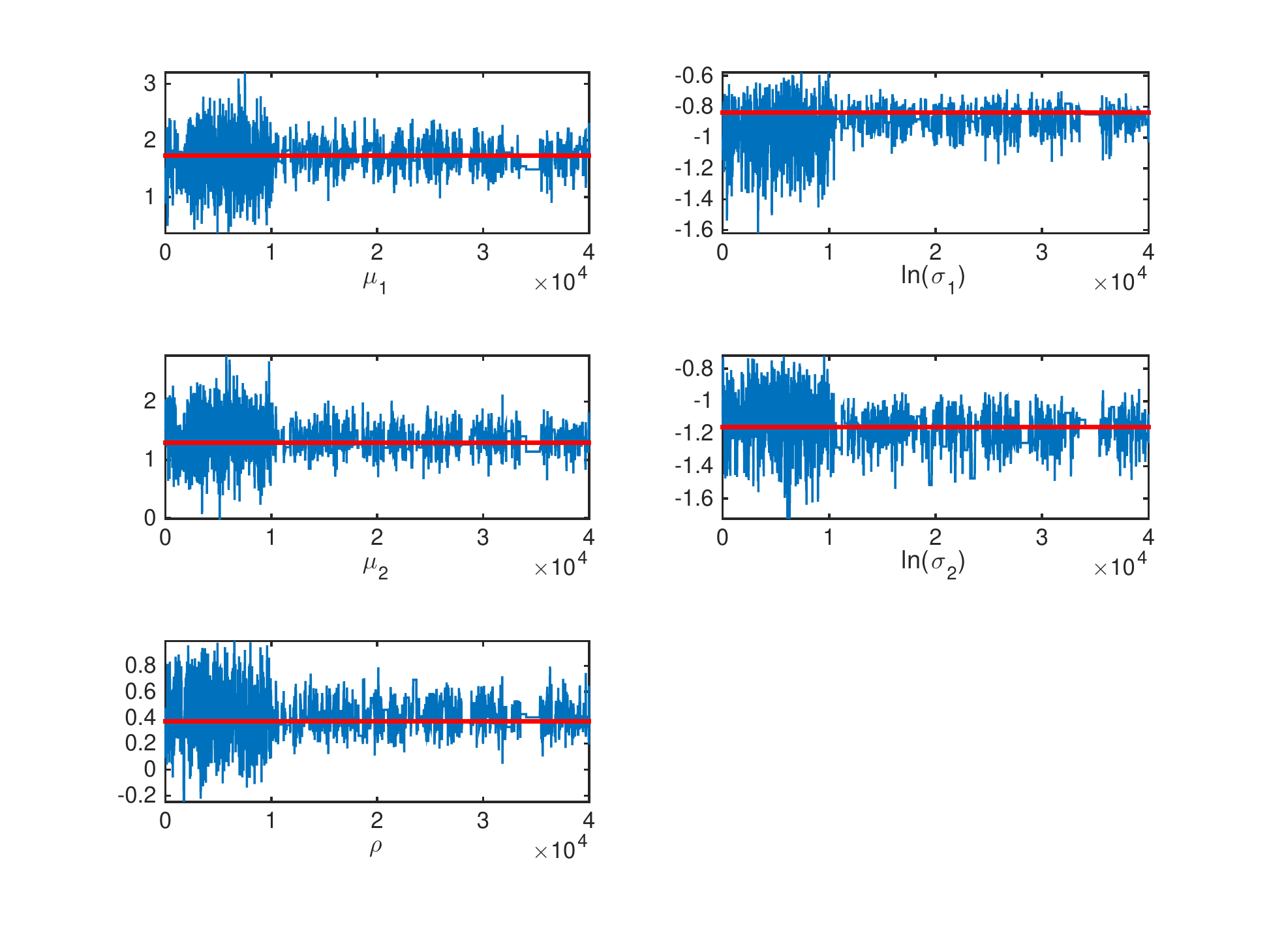}
\vspace{-1cm}
\caption{Geometric Brownian motion: (top) traceplots for $\mu_1$, $\log \sigma_1$; (middle) $\mu_2$, $\log\sigma_2$ and (bottom) $\rho$ when using ABC-DC and passing directly from 1 to 8 clones at iteration 10,000, without updating the threshold $\delta$. Horizontal lines are the exact MLEs.}
\label{fig:GBM_ABCDC_fast}
\end{figure}

For comparison, similar settings are applied to estimate the model parameters with ABC-DC.  The first 10,000 iterations are spent in the initial ABC-MCMC stage, where the threshold $\delta$ is updated once, at iteration 5,000, from 0.8 to 0.5.  At the end of the ABC-MCMC stage the acceptance rate is about 15\%. Then every 10,000 iterations the number of clones is increased progressively to 3, 5, 6, 7, and the final 40,000 iterations are run with $K=8$. In this last stage the acceptance ratio is 1\%. The simulation required 282 seconds, returning  $ \hat{\theta} = (1.7065 ,-0.8695 , 1.3254 , -1.1679  ,  0.4075 ) $. However, good estimates with ABC-DC  can be obtained with only 40,000 iterations, passing directly from 1 to 8 clones at iteration 10,000, without updating the threshold $\delta$, with a final acceptance rate of  3.3\%. In this case estimates are $ \hat{\theta} = ( 1.7145 , -0.8739 ,  1.3378 ,  -1.1804 , 0.4040)$, returned in only 112 seconds. Traceplots are shown in Figure \ref{fig:GBM_ABCDC_fast}.  A similar cut in the number of iterations failed with ABC-MCMC, since a slow reduction of the threshold $\delta$ was required in order to preserve an acceptable mixing of the chains. Results are summarized in Table \ref{tab:GBM-abcdc}.

The presented estimates are based on a single dataset generated from the set of parameters $\theta$, but in repetitions of the experiment ABC-DC has proved to be a robust method, as shown below. We now run $B=30$ independent simulations: for each simulation a different dataset is generated using the same value for the true parameter $\theta$ as considered in previous experiments. Then, each of these datasets is fitted using several algorithms, each algorithm returning the mean value of the proposed parameters. On the set of $B$ estimates $\hat{\theta}_{b}$   we compute the mean bias $\sum_{b=1}^{B}(\hat{\theta}_{b}-\theta)/B$ and the root mean square error (RMSE)  $\sqrt{\sum_{b=1}^{B}(\hat{\theta}_{b}-\theta)^2/B} $. Results obtained with ABC-MCMC and ABC-DC are comparable, see Table \ref{tab:GBM-robustness}. 

This example shows that both ABC-MCMC and ABC-DC can be applied with success to multi-dimensional SDE models when the likelihood function is unavailable, if we can identify a set of sufficient (or at least informative) summary statistics. The advantage of ABC-DC in this case is that we can obtain results comparably good in almost half of the time required for ABC-MCMC.

\begin{table}
\footnotesize
\centering{}

\begin{tabular}{cccccc}
\hline
\hline
 &  & 														& ABC-MCMC					&  ABC-DC \footnotesize{progressive}				&  ABC-DC \footnotesize{fast}	\\
&	True values		&	Exact MLE 	 & $K=1$, $\delta$=0.15 			& $K=8$, $\delta = 0.5$			& $K=8$, $\delta = 0.8$\\
&& & 100,000 iterations & 100,000 iterations &  40,000 iterations\\

\hline 
$\mu_{1}$		 	& 1.7  	&  1.7359  &     1.6426 							&	1.7065    	&   1.7145    \\
	
$\ln\sigma_{1}$  & -0.8 	& -0.8381  &     -0.8495 							&	-0.8695      &  -0.8739   \\

$\mu_{2}$ 		 	& 1.3		&  1.2913  &      1.2979 							&	1.3254	    &   1.3378    \\

$\ln\sigma_{2}$  & -1.2 	& -1.1599  &      -1.1599 							&   -1.1679    &     -1.1804  \\

$\rho$ 				 & 0.3  	&  0.3742 	&      0.3860  							&	  0.4075 	&   0.4040     \\
\hline
\end{tabular}
\caption {\footnotesize{True parameters, exact MLE, ABC-MCMC and ABC-DC estimates. Results comparably good can be obtained with ABC-DC in a bit more than half of the time required to ABC-MCMC. ``ABC-DC progressive'' considers the case where $\delta$ is reduced and $K$ is progressively increased. ``ABC-DC fast'' denotes a fixed $\delta$ and a $K$ increased rapidly.}}
\label{tab:GBM-abcdc}
\end{table}

\begin{table}
\footnotesize
\centering{}
\begin{tabular}{ccccccc}
\hline
\hline
   & \multicolumn{2}{c}{ABC-MCMC} & \multicolumn{2}{c}{ABC-DC \footnotesize{progressive}} & \multicolumn{2}{c}{ABC-DC \footnotesize{fast}}\\
 
 				  & Mean Bias & RMSE	&	Mean Bias & RMSE	&	Mean Bias &	RMSE \\
\hline 
$\mu_{1}$		&     -0.0911 & 0.2782 	 &  -0.0674 & 0.3605 	 & -0.0114 & 0.3021  \\
	
$\ln\sigma_{1}$ &     -0.0092 & 0.0408 	 &  -0.0134 & 0.0435 	 & -0.0409 & 0.0552  \\

$\mu_{2}$ 		&     0.0287  & 0.2309 	 & -0.0244 & 0.2241  	 &  0.0021 & 0.1727 \\
	
$\ln\sigma_{2}$ &     0.0109  & 0.0269 	 &  0.0083 & 0.0522   	& -0.0093 & 0.0317 \\

$\rho$ 		   	&    -0.0039  & 0.0423 	 & -0.0036 & 0.0561   	& 0.0321 & 0.0570  \\
\hline
\end{tabular}
\caption {\footnotesize{Mean bias and root mean square error for the parameters estimates on 30 different simulations.}}
\label{tab:GBM-robustness}
\end{table}

\section{Summary}

We have presented a strategy to integrate approximate Bayesian computation (ABC) in the so-called ``data cloning'' framework for approximate maximum likelihood estimation. A standard ABC-MCMC algorithm is initially used as a ``workhorse'' to locate an approximate maximum for the ABC posterior, which we use to center a Metropolis independent sampler, to be employed during the data-cloning stage. We note that the accuracy of the final inference, beyond mere identification of the location of the approximate MLE, can be enhanced for small $\delta$ and large $K$. However expecting our sampler to satisfy simultaneously both requirements is unrealistic as these are competitive criteria, particularly for highly erratic stochastic processes. That is, when considering an ABC framework it becomes increasingly unlikely to accept a proposed parameter as $K$ increases.
We note that for the considered examples ABC-DC produces reasonable inferences even though it is run with a larger $\delta$ than typically desired. In previous works using data-cloning MCMC (\citealp{lele-dennis-lutscher}, \citealp{baghishani-mohammadzadeh}, \citealp{jacquier-johannes-polson}) the algorithms were started at a high value of $K$, this opening the possibility for a chain to get stuck in some local maximum, instead we start the simulation with $K=1$ and then we increase it. Furthermore, in simulations considered in \cite{baghishani-mohammadzadeh} the starting parameter for their MCMC experiments was set to the exact MLE, which will \textit{of course} produce good results for $K$ large since their MCMC procedure starts in a peaked distribution already centred at its maximum. 

Besides statistical inference considerations, a scenario where we could see our method being employed is when considering a computationally expensive model simulator such that producing a single realizations from the model requires several seconds or minutes. Assume therefore that running many ($R$) iterations of ABC-MCMC is impractical, whereas running $\tilde{R}\ll R$ iterations of ABC-DC over $M>1$  processors is feasible (for simplicity, assume $K$ a multiple of $M$). Here the task of computing the cloned likelihood would be performed by distributing $K/M$ model simulations to each of the $M$ processors. 

A further application of the method can be envisaged when the experimenter is in need of reasonable and rapidly available estimates to be used as starting values for expensive procedures requiring a careful initialization, such as particle MCMC (pMCMC, \cite{andrieu2010particle}). For example in \cite{owen2015likelihood} it was found that using a specific ABC strategy (ABC-SMC) prior to starting pMCMC was of benefit, instead of investing a large amount of computational budget in trying to hand-tune pMCMC.

\section*{Acknowledgements} 
Research was funded by the Swedish Research Council (VR grant 2013-5167).


\bibliographystyle{elsarticle-harv} 
\bibliography{biblio}
\end{document}